\documentclass[namedreferences]{solarphysics}

\usepackage[optionalrh,solaromanenum,natbib,hyperref]{spr-sola-addons}
\usepackage{graphicx,pdflscape} 
\usepackage{color}    
\usepackage{breakurl} 
\hypersetup{colorlinks, citecolor=blue, filecolor=black, linkcolor=black, urlcolor=black}
\usepackage{rotating}

\makeatletter

\newcommand{\Rmnum}[1]{\expandafter\@slowromancap\romannumeral #1@}
\makeatother

\newcommand{\arcsec}{\hbox{$^{\prime\prime}$}}
\newcommand{\degree}{\hbox{$^{\circ}$}}
\newcommand{\eg}{{\it e.g.}}
\newcommand{\ie}{{\it i.e.}}

\begin{document}

\begin{article}

\begin{opening}

\title{Flaring Rates and the Evolution of Sunspot Group McIntosh Classifications}

%
 \author[addressref=aff1,corref,email={mccloska@tcd.ie}]{\inits{A.E. }\fnm{Aoife E. }\lnm{McCloskey}}
 \author[addressref=aff1,email={peter.gallagher@tcd.ie}]{\inits{P.T. }\fnm{Peter T. }\lnm{Gallagher}}
 \author[addressref=aff1,email={shaun.bloomfield@tcd.ie}]{\inits{D.S. }\fnm{D. Shaun }\lnm{Bloomfield}}

%
\runningauthor{A.E. McCloskey {\it et al.}}
\runningtitle{Flaring Rates and the Evolution of Sunspot Group McIntosh Classifications}

\address[id=aff1]{School of Physics, Trinity College Dublin, College Green, Dublin 2, Ireland}

\begin{abstract}
Sunspot groups are the main source of solar flares, with the energy to power them being supplied by magnetic-field evolution (\eg\ flux emergence or twisting/shearing). To date, few studies have investigated the statistical relation between sunspot-group evolution and flaring, with none considering evolution in the McIntosh classification scheme. Here we present a statistical analysis of sunspot groups from Solar Cycle 22, focusing on 24-hour changes in the three McIntosh classification components. Evolution-dependent $\geqslant$\,C1.0, $\geqslant$\,M1.0, and $\geqslant$\,X1.0 flaring rates are calculated, leading to the following results: (i) flaring rates become increasingly higher for greater degrees of upward evolution through the McIntosh classes, with the opposite found for downward evolution; (ii) the highest flaring rates are found for upward evolution from larger, more complex, classes (\eg\ Zurich D- and E-classes evolving upward to F-class produce $\geqslant$\,C1.0 rates of 2.66\,$\pm$\,0.28 and 2.31\,$\pm$\,0.09 flares per 24\,hours, respectively); (iii) increasingly complex classes give higher rates for all flare magnitudes, even when sunspot groups do not evolve over 24\,hours. These results support the hypothesis that injection of 
magnetic energy by flux emergence (\ie\ increasing in 
Zurich or compactness classes) leads to a higher frequency and magnitude of flaring.
\end{abstract}

%

\end{opening}

%
\section{Introduction}\label{sec:intro}

Solar flares are known to originate in active regions on the Sun, and they are the result of the rapid release of large quantities of energy \citep[up to 10$^{32}$ ergs;][]{Emslie2012} from complex magnetic-field structures rooted in their sunspot groups. This release of magnetic energy can lead to the acceleration of highly energetic particles and emission of high-energy radiation that can affect the performance and reliability of technology in the near-Earth space environment (\ie\ space weather).
Timescales for space-weather events, from first detection to arrival in the near-Earth environment, range from instantaneous (solar flare electromagnetic radiation) to tens of minutes (solar energetic particles) or hours \citep[coronal mass ejections;][]{Vrsnak2013}.
There is a great need to develop a better understanding of the conditions that lead to the production of solar flares, due to the simultaneous nature of their initial detection and Earth impact.

Historically, the complexity of sunspot groups has been investigated as an indicator of potential flaring activity. A classification scheme describing their magnetic complexity was established by \cite{Hale1919} and is known as the Mount Wilson classification scheme. It originally consisted of three parameters to describe the mixing of magnetic polarities in sunspot groups: $\alpha$ (unipolar); $\beta$ (bipolar); $\gamma$ (multipolar). Early work relating these magnetic classifications to flare productivity showed that sunspot groups of increasingly complex Mount Wilson class (\eg\ $\beta$ to $\beta\gamma$ to $\gamma$) were found to produce increasing frequencies of flaring \citep{Giovanelli1939}. This scheme was later extended to include close ($\leqslant$\,2\degree) mixing of umbral magnetic polarities within one penumbra, known as the $\delta$-configuration \citep{Kunzel1960}. Studies including this extended scheme have shown that groups that achieve greater magnetic complexity (\eg\ $\beta\gamma\delta$-configuration) and larger sunspot area (a proxy for total magnetic flux) produce flares of greater magnitude at some point in their lifetime \citep{Sammis2000}.

Analogously, a classification scheme describing the white-light structure of sunspot groups was developed, originally by \cite{Cortie1901} and later modified and expanded to include a wider range of parameters \citep{Waldmeier1947,McIntosh1990}. Currently this is referred to as the McIntosh scheme, which always consists of three components [Zpc]:
\begin{description}
\item[Z] modified Zurich class, describing longitudinal extent of the sunspot group;
\item[p] penumbral class, describing size/symmetry of largest sunspot's penumbra;
\item[c] compactness class, describing interior spot distribution of the group.
\end{description}
As these classifications are the primary focus of this work, Section~\ref{sssec:mcint} describes the individual McIntosh components in greater detail.

Statistical analysis has previously been carried out on sunspot-group McIntosh classifications to produce historically averaged rates of flaring. Similar to the magnetic-complexity work of \cite{Giovanelli1939}, it was found that sunspot groups with higher McIntosh structural complexity classes (corresponding to larger extent, large and asymmetric penumbrae, and more internal spots) produced higher flaring rates overall \citep{McIntosh1990,Bornmann1994}.

In more recent years, several studies have investigated magnetic properties of sunspot groups that are thought to play an important role in flare production. It has been shown that flares most commonly occur in regions that display rapidly emerging flux \citep{Schmieder1994} or twisted, non-potential magnetic fields, a signature of stored free magnetic energy \citep{Hahn2005}. Examples of derived, point-in-time magnetic properties include strong horizontal gradients of magnetic field close to polarity inversion lines -- the $R$-value of \citet{Schrijver2007} and the $^{L}$WL$_{\mathrm{SG}}$ value of \citet{Falconer2008} -- and large effective connected magnetic field \citep[$B_{\mathrm{eff}}$;][]{Georgoulis2007}. These derived properties all show a potential for use in flare forecasting through varying degrees of correlation with flaring activity. However, there has yet to be any large-scale statistical analysis on applying these properties to forecast flares.

Such large-scale statistical analyses have been carried out on historical records of sunspot properties and their relation to flaring activity. \cite{Gallagher2002} implemented a flare-forecasting method using historical McIntosh classifications to produce flare probabilities from average flare rates under the assumption of Poisson statistics. Although only taking into account morphological properties, the McIntosh--Poisson forecasting method has comparable levels of success to other much more complex techniques \citep{Bloomfield2012} and expert-based systems \citep[\eg][]{Crown2012,Bloomfield2016}.

However, none of the works described so far take into account a key factor in pre-flare conditions, namely the evolution of the sunspot-group properties. The energy that is available for flaring is governed by the Poynting flux through the solar surface, which can be modified by changes in total magnetic flux (through emergence or submergence) and reorientation of the magnetic field (through twisting, shearing, or tilting). 
 In terms of flux emergence, \citet{Schrijver2005} found that active-region non-potentiality (correlated with higher likelihood of flaring) was enhanced by flux emergence in the 10\,--\,30\,hours prior to flares. In addition, \citet{Lee2012} studied the most flare-productive McIntosh classifications and 24-hour changes in their sunspot group area (\ie\ decreasing, steady, or increasing), finding a noticeable increase in flare-occurrence rates for sunspot groups of increasing area. Regarding the reorientation of the magnetic field, \citet{Murray2012} found that local concentrations of magnetic flux at flare locations displayed a field-vector inclination ramp-up towards the vertical before flaring. 
In addition, this reorientation of the field resulted in a corresponding pre-flare increase in free magnetic energy that then decreased after the flare \citep{Murray2013}.
 These works highlight that sunspot-group property evolution is an important indicator of flaring activity. However, there has not yet been a study of the temporal evolution of sunspot-group classifications and its potential for use in flare forecasting.

Here we present statistical analysis of the evolution of sunspot groups in terms of their McIntosh white-light structural classifications and associated flaring rates. In Section \ref{sec:data_anal}, the distribution of sunspot groups across McIntosh classes is presented along with evidence for mis-classification in limb regions. Section~\ref{sec:results} then focuses on the main results and discussion of the study. Analysis of the overall evolution of sunspot-group McIntosh classes is presented in Section~\ref{ssec:class_evol}, with the class-specific evolution discussed in Section~\ref{ssec:zurich_class_evol}. Flaring rates associated with these class-specific evolution steps are included in Section~\ref{ssec:flare_rates} for three different flaring levels (\ie\ $\geqslant$\,C1.0, $\geqslant$\,M1.0, and $\geqslant$\,X1.0), while our conclusions and future direction are presented in Section~\ref{sec:conc}.

\section{Data and Analysis}\label{sec:data_anal}

\subsection{Data Sources}
The data studied in this article were obtained from historical catalogs of sunspot classifications and properties that were collected by the National Oceanographic and Atmospheric Administration (NOAA) Space Weather Prediction Center (SWPC). 
The aim of this work is to determine flaring rates associated with McIntosh class evolution for future use in flare forecasting. In order to ensure an independent data set to test on more recent data in future studies (\eg\ Solar Cycles 23 and 24) we have chosen to limit our analysis in this article to data from 
Solar Cycle 22 over the period 1 December 1988 to 31 July 1996, inclusive.

These NOAA data are collected by a world-wide network of ground-based optical telescopes (Solar Observing Optical Network) that provide daily reports of sunspot observations and their properties. These data are then collated by SWPC and published as the Solar Region Summary (SRS) each day at 00:30 UT. Published properties include the NOAA active-region number, heliographic coordinates, McIntosh and Mount Wilson classifications, longitudinal extent, and sunspot-group area. In addition, region-associated solar flares were obtained from data collected by the \emph{Geostationary Operational Environmental Satellite} (GOES). Flare data used in this study includes all GOES 1\,--\,8\,\AA\ soft X-ray flares of C-class and above (\ie\ $\geqslant$\,$10^{-6}$\,Wm$^{-2}$). The NOAA/SWPC SRS and GOES event list data from 1996 onward is publicly available from the NOAA/SWPC online archive (\url{ftp://ftp.swpc.noaa.gov/pub/warehouse/}). However, it is noted that the data analyzed in this article cover 1 December 1988 to 31 July 1996, inclusive. Data from this time period are not publicly available and were instead obtained directly from NOAA/SWPC staff (C.C. Balch 2011, private communication).

Using this historical data, McIntosh classifications were extracted for each unique spotted NOAA active-region entry. Each entry corresponds to the classification of a region on a 24-hour basis, meaning that each unique region can have several individual entries over its lifetime. This yielded a total of  
18,736 entries from 1 December 1988 to 31 July 1996, inclusive, corresponding to 2708 unique NOAA regions. A total of 7648 GOES soft X-ray flares were recorded over the same period, comprising of 6149 C-class, 1383 M-class, and 116 X-class.

\subsection{McIntosh Sunspot Group Classifications}\label{sssec:mcint}

The McIntosh classification scheme describes the white-light structure of sunspot groups and is composed of 60 allowed classification combinations derived from 17 different parameters \citep{McIntosh1990}. As introduced in Section~\ref{sec:intro}, there are three components to each unique McIntosh classification (\ie\ Zpc), namely the modified Zurich (Z), penumbral (p), and compactness (c) classes.

\begin{table}[!h]
\caption{Parameters and characteristics describing the modified Zurich classes of the McIntosh classification scheme}
\label{T-Zurich}
\begin{tabular}{cccccc}     
\hline                   
Zurich & Unipolar    & Bipolar   & Penumbra   & \multicolumn{2}{c}{Largest spot separation in group} \\
class  & group       & group     & present    & \multicolumn{2}{c}{[Heliographic degrees]} \\
       &             &           &            & Minimum & Maximum \\
\hline
A      & \checkmark & $\times$   & $\times$   & $\cdots$  & 3\degree  \\ 
B      & $\times$   & \checkmark & $\times$   & 3\degree  & $\cdots$  \\ 
H      & \checkmark & $\times$   & \checkmark & $\cdots$  & 3\degree  \\
C      & $\times$   & \checkmark & \checkmark \tabnote{Penumbra only present on either leading or trailing sunspot} & 3\degree & $\cdots$  \\
D      & $\times$   & \checkmark & \checkmark & 3\degree  & 10\degree \\
E      & $\times$   & \checkmark & \checkmark & 10\degree & 15\degree \\
F      & $\times$   & \checkmark & \checkmark & 15\degree & $\cdots$  \\
\hline
\end{tabular}
\end{table}

\begin{table}[!h]
\caption{Parameters and characteristics describing the penumbral classes of the McIntosh classification scheme}
\label{T-Pen}
\begin{tabular}{cccc}     
\hline                   
Penumbral & Penumbra & North--south diameter   & Symmetric\\
class     & present  & [Heliographic degrees] & penumbra\\
\hline
X & $\times$   & $\cdots$                & $\cdots$  \\ 
R & \checkmark & $\cdots$ \tabnote{Penumbra extends $<$\,3\arcsec\ from umbra} & $\cdots$\\ 
S & \checkmark & $\leqslant$\,2.5\degree & \checkmark  \\
A & \checkmark & $\leqslant$\,2.5\degree & $\times$   \\
H & \checkmark & $>$\,2.5\degree         & \checkmark  \\
K & \checkmark & $>$\,2.5\degree         & $\times$  \\
\hline
\end{tabular}
\end{table}

\begin{table}[!h]
\caption{Parameters and characteristics describing the compactness classes of the McIntosh classification scheme}
\label{T-Comp}
\begin{tabular}{cccc}     
\hline                   
Compactness & Unipolar   & Bipolar    & Interior spot\\
class       & group      & group      & distribution\\
\hline
X           & \checkmark & $\times$   & $\cdots$ \tabnote{Unipolar groups are treated as having no interior}\\
O           & $\times$   & \checkmark & Open \tabnote{Few or no internal spots present}\\ 
I           & $\times$   & \checkmark & Intermediate\\
C           & $\times$   & \checkmark & Compact\\
\hline
\end{tabular}
\end{table}

The Zurich classes describe the large-scale structure of sunspot groups in terms of the parameters given in Table~\ref{T-Zurich}. An important property of the Zurich classification is that it provides information on the longitudinal extent of the sunspot group. This essentially acts as a proxy for the total amount of magnetic flux, both in terms of its emergence and its disappearance through submergence. For example, a group evolving from Zurich A- to B-class indicates growth from unipolar to bipolar (\ie\ flux emergence). Evolution from A- to H-class indicates development of a penumbra in the primary spot with no limit on the longitudinal extent of the group. Additionally, evolution from B-class to the D-/E-/F-classes corresponds to the development of penumbrae on both leading and trailing sunspots with a probable increase in longitudinal extent.\footnote{Note, there is no upper limit to the longitudinal extent of a B-class sunspot group indicated in Table~\ref{T-Zurich}, but these groups are typically of smaller extent.} The evolution of magnetic flux is an important indicator of potential flaring activity, captured in this work by the evolution in Zurich class.

In terms of smaller-scale structure, the McIntosh penumbral classes provide information on the size and symmetry of the penumbra of the largest sunspot in the group (Table~\ref{T-Pen}). The overall symmetry of sunspot penumbrae can be interpreted as an indicator of the magnetic field topology, such as the degree of twisting and shearing present in the sunspot group. Although the modified Zurich classes indicate the overall extent of sunspot groups, the filling of the space between the leading and trailing spots is described by the McIntosh compactness class (Table~\ref{T-Comp}). Therefore, this classification component can also be an indicator of total magnetic flux, similar to the modified Zurich class.

It is worth briefly considering the evolutionary paths that sunspot groups take through the McIntosh class components. The lower-ordered Zurich classes of A, B, H, and C differ in multiple characteristics (\ie\ uni-/bi- polarity, presence of penumbrae on one or both ends, longitudinal extent of group), while the higher-ordered Zurich classes D, E, and F only differ in longitudinal extent. This results in sunspot groups being capable of evolving by non-incremental steps in Zurich class (\ie\ one or two classes may be skipped during the early growth phase of a group). McIntosh penumbral class is also capable of non-incremental evolution, as steps of +\,1 can indicate changes from symmetric to asymmetric penumbrae only (\eg\ S to A, or H to K) while evolution steps of +\,2 can indicate growth in penumbral size only (\eg\ S to H, or A to K). In contrast, the interior spot distribution of sunspot groups will more frequently evolve by incremental steps in McIntosh compactness class, with increasing from no internal spots (\ie\ X or O, depending on uni-/bi polarity of group) to some spots (\ie\ I) to nearly continuous spots between the primary leading and trailing spots (\ie\ C).

The combination of the three McIntosh class components has been shown to capture differences in the rates of flare production from sunspot groups \citep[see, \eg,][]{Bornmann1994}, and should therefore be an appropriate system to use in our evolution-dependent analysis. We include only the Zurich-class results in the main text of this article because of the large quantity of plots. Figures arising from the equivalent analysis of the penumbral and compactness classes are available in Appendices~\ref{app_pen} and \ref{app_comp}, respectively.

\begin{figure}[!ht]
\begin{center}
\includegraphics[width=0.95\textwidth]{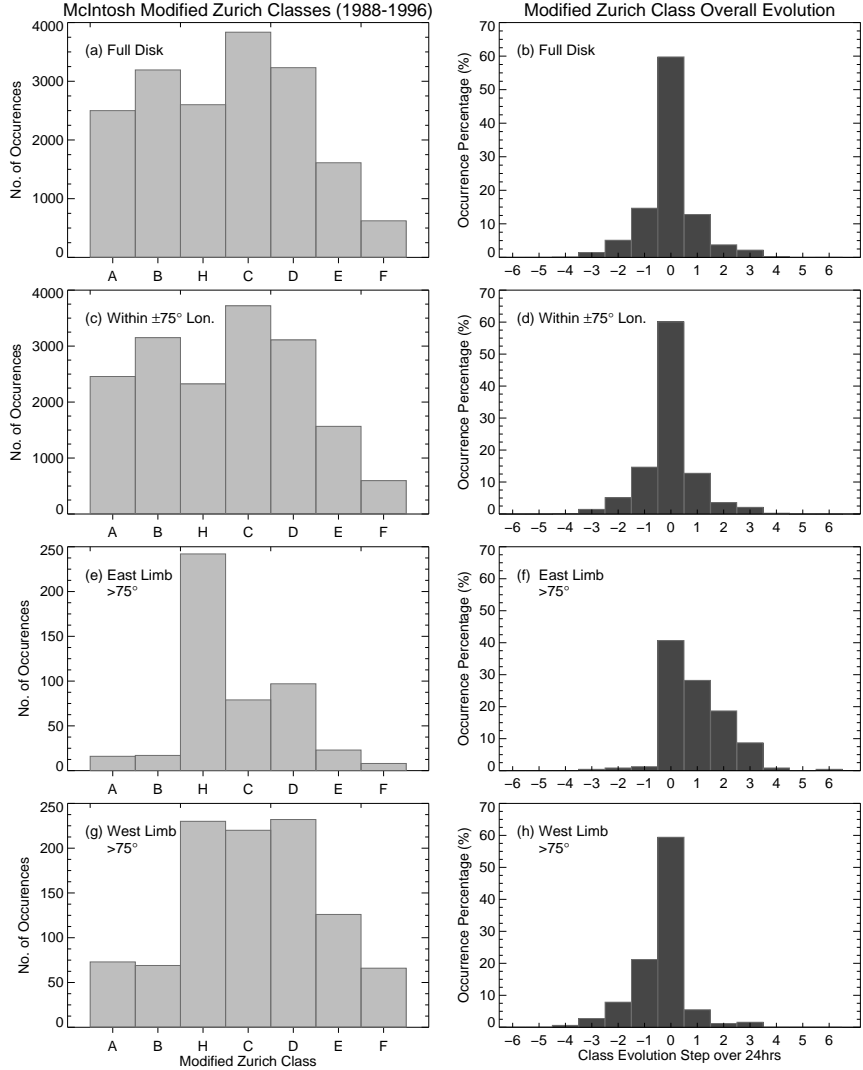}
\caption{Frequency histograms of each modified Zurich class in the McIntosh classification scheme (left column) and occurrence percentage histograms of their overall evolution steps on 24-h timescales (right column). Each row presents data from different spatial locations on the Sun: full disk (panels a\,--\,b); within $\pm$\,75\degree\ Heliographic longitude (panels c\,--\,d); east limb (panels e\,--\,f); west limb (panels g\,--\,h). Positive evolution steps correspond to stepping downwards through the Zurich classes in Table~\ref{T-Zurich}.}
\label{fig:hist_freq}
\end{center}
\end{figure}

\subsection{Modified Zurich Class Occurrences}\label{ssec:class_occ}

The frequency distribution of all modified Zurich classes that were observed from 1 December 1988 to 31 July 1996, inclusive, is provided in Figure~\ref{fig:hist_freq}. The full-disk distribution in Figure~\ref{fig:hist_freq}a shows a high frequency of classes A\,--\,D, with the most frequently observed being C-class (\ie\ bipolar sunspot groups with penumbra on only one of either the leading or trailing spots). Zurich E- and F-classes are much less frequently observed, comprising only 12\,\% of all observations. These represent the largest, most complex sunspot groups that are most often associated with the production of large-magnitude flares. The relative occurrence frequency of Zurich classes agrees well with previous statistical analysis of other periods, such as the \cite{McIntosh1990} study of 1969\,--\,1976 in Solar Cycle 20.

To consider the effect that viewing angle (and hence foreshortening) may have on the classification of sunspot groups, we split the distribution of Figure~\ref{fig:hist_freq}a into three spatial regimes: within $\pm$\,75\degree\ Heliographic longitude (Figure~\ref{fig:hist_freq}c); east limb (Figure~\ref{fig:hist_freq}e); west limb (Figure~\ref{fig:hist_freq}g). The 75\degree\ Heliographic longitude cut-off point for limb regions was chosen based on the 24-hour issuing time scale of the NOAA/SWPC sunspot group classifications. Coupled with a typical solar rotation rate of $\approx$\,14\degree\ per day \citep[for 30\degree\ latitude;][]{Snodgrass1990}, the choice of 75\degree\ ensures one classification in the east or west limb regions for sunspot groups rotating over either limb. This allows for the potential mis-classification of a sunspot group on its first or last day to be excluded from the central disk portion (\ie\ within $\pm$\,75\degree) and examined separately.

The distribution of Zurich classes within $\pm$\,75\degree\ of central meridian is essentially similar to that of the full disk, while those at the east and west limbs show a clear divergence. The east limb has a distinctive deficit of A- and B-classes and dominance of H-class (\ie\ unipolar with penumbra), indicating an over-classification of H-class groups. This is most likely due to limb foreshortening that limits the accuracy of measuring the longitudinal extent and uni-/bi- polarity of sunspot groups as they rotate into or out of view. For example, when large Zurich D-, E-, or F-class groups rotate over the east limb, their mature leading spots can be misinterpreted as unipolar H-class groups. The west-limb region also differs from the within $\pm$\,75\degree\ case, but with reduced magnitude to that of the east limb. As a result of this apparent mis-classification, we divide the data into within $\pm$\,75\degree\ Heliographic longitude, east limb, and west limb for all subsequent analysis. Equivalent versions of Figure~\ref{fig:hist_freq} are presented for the penumbral and compactness classes in Appendices~\ref{app_pen} and ~\ref{app_comp}, respectively.

\section{Results and Discussion}\label{sec:results}
In this section, we present an in-depth analysis of sunspot-group evolution in terms of the McIntosh modified Zurich classes. Firstly, the evolution of all Zurich classes are analyzed in Section~\ref{ssec:class_evol} irrespective of their starting/ending class, while in Section~\ref{ssec:zurich_class_evol} these are broken down into specific evolutionary steps through the Zurich classes. Finally, Section~\ref{ssec:flare_rates} presents flaring rates that result from these evolutionary steps along with their uncertainties. We remind the reader that equivalent figures resulting from applying this analysis to the penumbral and compactness classes are given in Appendices~\ref{app_pen} and \ref{app_comp}, respectively.

\subsection{Overall Evolution in Zurich Class}\label{ssec:class_evol}

The overall evolution of Zurich classes was examined in terms of upward or downward steps over 24-hour periods -- this timescale was chosen as the NOAA/SWPC SRS data are published once daily. The right-hand column of Figure~\ref{fig:hist_freq} presents these 24-hour changes through Zurich class in numeric form, whereby a change of +\,1 in evolution space indicates an increase in Zurich complexity by one step downwards through Table~\ref{T-Zurich} (\eg\ B- to C-class, C- to D-class).

Individual panels in the right-hand column of Figure~\ref{fig:hist_freq} again cover different spatial locations on the Sun. Focusing on the full-disk data in Figure~\ref{fig:hist_freq}b, the distribution is dominated by zero evolution in Zurich class while also displaying a slight asymmetry with positive skew. This skew (referring to a higher percentage of downward evolution steps) can be explained by sunspot groups having a longer decay phase than flux-emergence phase during their lifetime. Excluding limb regions, the evolution of sunspot groups within $\pm$\,75\degree\ Heliographic longitude is shown in Figure~\ref{fig:hist_freq}d. This is qualitatively similar to Figure~\ref{fig:hist_freq}b, unsurprisingly as it dominates the spatial extent of the full disk.

Sunspot groups close to the east and west limbs (\ie\ $\geqslant$\,75\degree\ 
longitude) were examined separately and their results given in Figures~\ref{fig:hist_freq}f and \ref{fig:hist_freq}h, respectively. Both distributions show stronger asymmetry of evolution steps compared to sunspot groups within $\pm$\,75\degree\ Heliographic longitude, but they also display oppositely biased behaviour to one another. There is a strong bias for upward evolution in Zurich class at the east limb, whereas downward evolution dominates at the west limb. This reinforces the argument that the majority of sunspot groups near the limbs are significantly affected by foreshortening, whereby a limited extent of a group is visible at the start (east limb) or end (west limb) of its disk transit. This produces evolution-frequency distributions at the limbs that do not represent the true evolution of sunspot groups, strengthening the justification for removing limb regions for our calculation of flaring rates in Section~\ref{ssec:flare_rates}.

The overall evolution distributions for McIntosh penumbral and compactness classes show qualitatively similar results to that of the Zurich-class analysis 
(see Appendices~\ref{app_pen} and \ref{app_comp}, respectively). It can be seen that the least complex classes in both penumbral (\eg\ X, S, and A) and compactness (\eg\ X and O) are 
most frequently observed. Additionally, the mis-classification of groups at the limbs is again evidenced by oppositely biased, asymmetric distributions of evolution steps at the east and west limbs.

\begin{figure}[!ht]
\begin{center}
\includegraphics[width=0.95\textwidth]{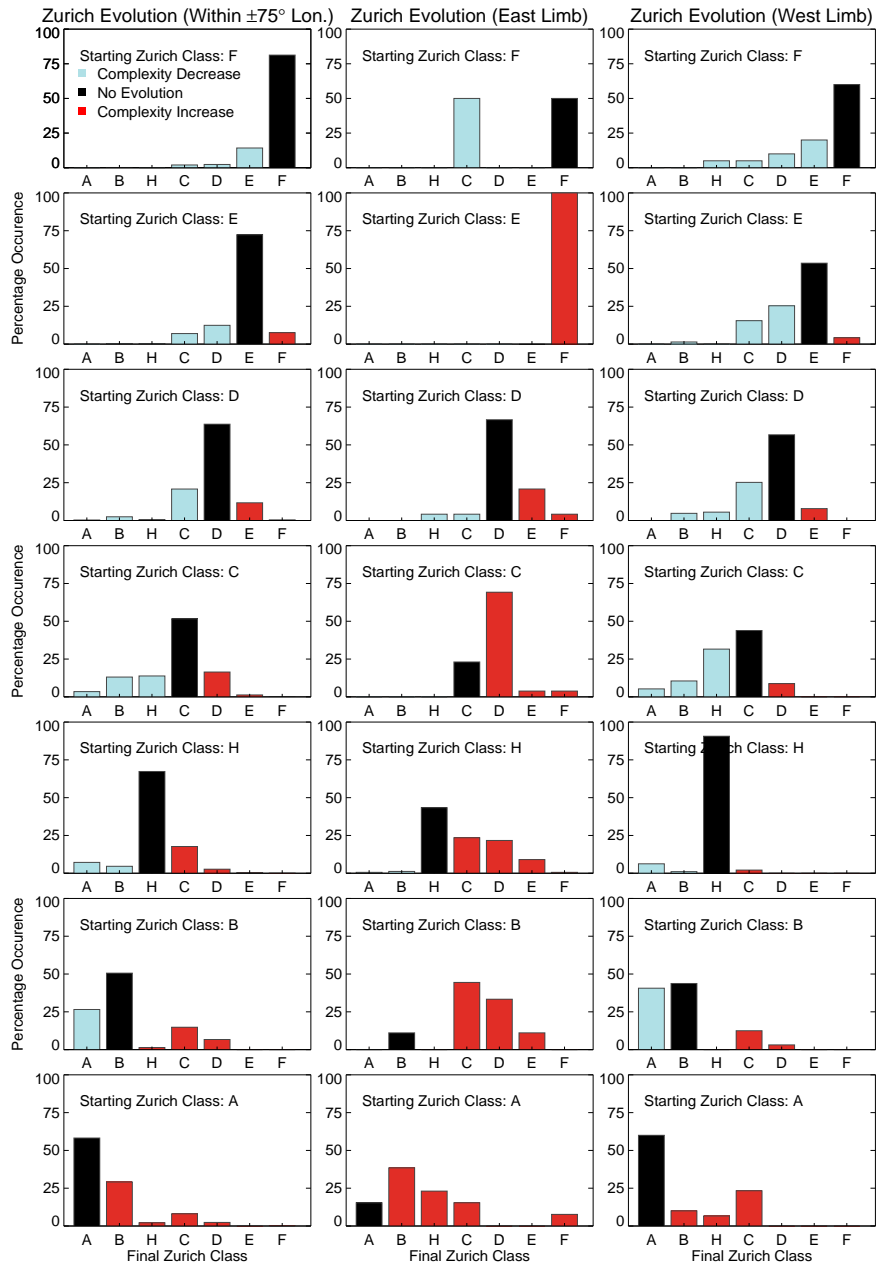}
\caption{Zurich class 24-hour evolution histograms. Each column concerns different locations on the Sun: within $\pm$\,75\degree\ longitude (left); east limb (centre); west limb (right). Each row presents evolution from a different starting class, while bars give the percentage of that starting class coloured by evolution: no change (black); upward evolution (red); downward evolution (blue).}
\label{fig:zur_evol}
\end{center}
\end{figure}

\subsection{Zurich Class-Specific Evolution}\label{ssec:zurich_class_evol}

The evolution of McIntosh classifications was examined in detail to determine where each sunspot group evolves from and to in terms of Zurich, penumbral, and compactness classes, with the eventual intention of calculating a corresponding flaring rate for each specific class-evolution step. Evolution in only one McIntosh classification component was considered for this analysis to attempt to keep the flaring rates of Section~\ref{ssec:flare_rates} statistically significant.

Frequency distributions of evolution occurrence through Zurich class are given in Figure~\ref{fig:zur_evol}, with the three columns presenting separate locations on the Sun: within $\pm$\,75\degree\ Heliographic longitude (left column); east limb (centre column); west limb (right column). Each is divided into seven panels, corresponding to all possible Zurich classes (see Table~\ref{T-Zurich}) observed before a sunspot group evolves over 24\,hours. In each panel, histogram bars represent the evolution of individually tracked sunspot groups in terms of their starting and final Zurich classes over 24\,hours. These are given as a percentage of the total number of occurrences for that starting Zurich class in order to visualize evolution from Zurich classes that have low occurrence, with the explicit values for the within $\pm$\,75\degree\ case (left column) provided in Appendix~\ref{app_zur}. Furthermore, the histogram bars are coloured to represent their form of evolution in that panel: increase in complexity (red); no change in complexity (black); decrease in complexity (blue). 
 Taking the left-bottom panel of Figure~\ref{fig:zur_evol} as an example, ``Starting Zurich class: A'' shows that $\approx$\,60\,\% of Zurich A-class sunspot groups do not evolve on a 24-hour timescale, while the second-most frequent evolution is for A-class to evolve upwards in complexity to B-class (\ie\ becoming bipolar, but maintaining no penumbra).

If we first consider the evolution of sunspot groups that are not located near the east and west limbs (Figure~\ref{fig:zur_evol} left column, within $\pm$\,75\degree\ Heliographic longitude), in all starting-class panels the majority of sunspot groups remain as the same Zurich class rather than evolving upward or downward in complexity (corresponding to the ``0'' evolution step in the overall evolution analysis of Figure~\ref{fig:hist_freq} in Section~\ref{ssec:class_evol}). For the ``Starting Class: C'' panel, it is notable that if a sunspot group does change Zurich class it is almost equally likely to evolve upward (to D-class) as it is to evolve downward (to H-class), with all transitions available for evolution. This indicates that semi-mature sunspot groups (\ie\ of intermediate size with penumbra present on only one end) are equally likely to emerge flux and form additional penumbrae as they are to decay.


Interestingly, for ``Starting Class: H'' there are two dominant evolutionary steps, namely no evolution or for a H-class to evolve into C-class. This transition is again constrained by the definition of H-class (\ie\ unipolar with penumbra) and C-class (\ie\ bipolar with penumbra on one end only). Hence, this evolution corresponds to the emergence of opposite-polarity spots without penumbra into the sunspot group. It is worth noting that H-class groups were originally thought to represent the final stages of a sunspot group life-cycle, when flux has nearly fully decayed. However, this analysis shows that these sunspot groups frequently emerge magnetic flux than to undergo decay.

It is also notable that the largest classes (\ie\ E and F) are seldom observed to evolve significantly in terms of Zurich class (\ie\ typically $\pm$\,1 evolution step). This indicates that when sunspot groups are large ($>$\,10\degree) 
 they do not decay rapidly (\eg\ becoming unipolar) on a 24-hour timescale. Additionally, there seems to be a preference for evolution by $\approx$\,one step in Zurich class present throughout all of the starting-class evolution panels of Figure~\ref{fig:zur_evol}, indicating that rapid evolution of a sunspot group is extremely unlikely over 24\,hours.

In contrast, sunspot groups close to the east and west limbs (Figure~\ref{fig:zur_evol} centre and right columns, respectively) again display significant and opposite trends in terms of their Zurich-class evolution. Sunspot groups are dominated by evolution upward in Zurich class when close to the east limb, with downward evolution essentially missing for all but those sunspot groups that start as D- and F-class. Similar but opposite behaviour is observed for the evolution of groups close to the west limb, with a significantly higher percentages evolving downward. These systematically aberrant behaviours close to the limbs strengthen the argument for removing limb regions and considering only the evolution of sunspot groups within $\pm$\,75\degree\ longitude to determine evolution-dependent flaring rates.
 
Equivalent analyses of penumbral and compactness class-specific evolution are presented in Appendices~\ref{app_pen} and \ref{app_comp}, respectively. These distributions of class-specific evolution occurrence show qualitatively similar behaviour to that of the Zurich classes within $\pm$\,75\degree\ Heliographic longitude (Figure~\ref{fig:zur_evol}, left column), with no evolution on a 24-hour timescale again dominating each starting-class panel.

\subsection{Zurich Evolution-Dependent Flaring Rates}\label{ssec:flare_rates}

In the context of this study, flaring rates are calculated as the average number of flares produced in the 24\,hours following an evolution in the McIntosh class of a sunspot group. For example, groups starting as Zurich C-class and evolving into D-class have a total of 562 occurrences and produced a total of 370 flares above C1.0 in GOES magnitude. Therefore, the $\geqslant$\,C1.0 flaring rate for C- to D-class evolution is 0.66\,$\pm$\,0.04 flares per 24\,hours.

\begin{figure}[!ht]
\begin{center}
\includegraphics[width=0.95\textwidth]{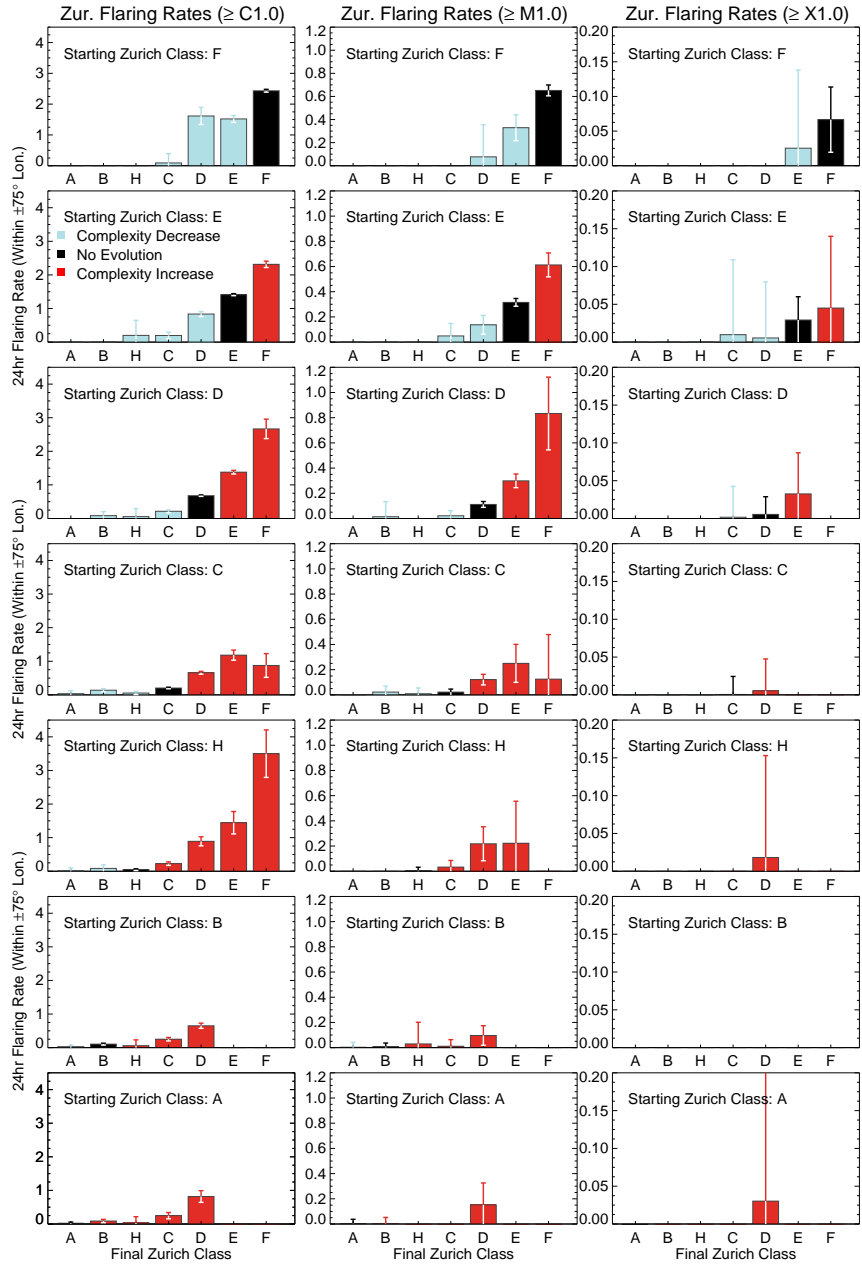}
\caption{Zurich evolution-dependent 24-hour flaring rates from groups within $\pm$\,75\degree\ longitude. Each column concerns different flaring levels: $\geqslant$\,C1.0 (left); $\geqslant$\,M1.0 (centre); $\geqslant$\,X1.0 (right). As in Figure~\ref{fig:zur_evol}, each row shows evolution from a different starting class and histogram bars are coloured by evolution: no change (black); upward evolution (red); downward evolution (blue).}
\label{fig:zur_flare_nolimb}
\end{center}
\end{figure}

Figure~\ref{fig:zur_flare_nolimb} displays the flaring rates for the specific evolutionary steps in Zurich class that were shown previously in the percentage occurrence plots of Figure~\ref{fig:zur_evol}, with values provided in Appendix~\ref{app_zur}. These plots depict flaring rates calculated only from those sunspot groups within $\pm$\,75\degree\ longitude (\ie\ excluding the limb regions), with the three columns now presenting rates at different flaring levels: $\geqslant$\,C1.0 (left column); $\geqslant$\,M1.0 (centre column); $\geqslant$\,X1.0 (right column). In contrast to their occurrence-frequency counterparts, these distributions do not show the highest rates of flaring for sunspot groups that do not evolve. Taking ``Starting Zurich Class: D'' as an example, there is an increasingly higher rate of $\geqslant$\,C1.0 flaring for evolution upward to more complex Zurich classes. This behaviour exists for basically all starting classes in the left column of Figure~\ref{fig:zur_flare_nolimb} -- the greater the degree of upward evolution in Zurich class, the higher the rate of flaring -- with the opposite behaviour (\ie\ sequentially lower rates) for greater degrees of downward evolution. It is worth noting that this behaviour is still present (albeit somewhat less pronounced) for the higher flare-magnitude cases of $\geqslant$\,M1.0 and $\geqslant$\,X1.0 in the centre and right columns of Figure~\ref{fig:zur_flare_nolimb}, respectively.

It is notable that the flaring rates for each no-evolution case (black bars) increases with increasingly complex Zurich class -- \eg\ sunspot groups starting as D-class that do not evolve produce less flares than groups starting as E-class with no evolution (Figure~\ref{fig:zur_flare_nolimb}, second and third panels from top). This indicates that even when there is no significant level of flux emergence or decay in a sunspot group, the magnitude of the flaring rate scales with the Zurich complexity.

In addition, sunspot groups starting as Zurich H-class produce a significant increase in flaring rate with increasing upward evolution compared to all of the other starting classes. The evolution step that produces the greatest $\geqslant$\,C1.0 flaring rate is for groups starting as H-class and evolving to F-class, even though this is one of the least-frequently observed evolution steps to occur across all of the observed evolutionary steps (Figure~\ref{fig:zur_evol}). The upward evolution from H-class to D-, E-, and F-class are a physical manifestation of rapid flux emergence in these sunspot groups (as the possibility of limb foreshortening has been removed from this sample). For example, the evolution from H-class (\ie\ unipolar with penumbra) into F-class (\ie\ bipolar and large extent, with penumbrae on both leading and trailing spots) indicates that a large amount of magnetic flux has emerged within the sunspot group over 24\,hours, and this leads to a high rate of flare production. The majority of these are GOES C1.0\,--\,C9.9 flares, as evidenced by the low $\geqslant$\,M1.0 and $\geqslant$\,X1.0 flaring rates. However, if a sunspot group starts as D- or E-class (\ie\ bipolar and moderate/large extent) and evolves to F-class (\ie\ bipolar and largest extent), it is seen to produce greater flaring rates of large-magnitude flares (\ie\ M- and X-class) relative to an evolution from H- to F-class. This indicates there may be an upper limit to the magnitude of flares produced by the evolution of a sunspot group that is dependent on the starting Zurich class. In other words, smaller and less-complex sunspot groups (at least in terms of Zurich class) are not observed to produce significant numbers of large-magnitude flares, independent of their evolution, while large and complex sunspot groups produce the majority of M- and X-class flares.

To quantify the significance of these flaring rates we also calculated their associated uncertainties. It has been shown that the distribution of flares can be approximated quite accurately by that of a Poisson distribution \citep{Gallagher2002}. This has proven to be a successful approximation in both a purely statistical examination \citep{Wheatland2001} and in implementing flare-forecasting methods \citep{Bloomfield2012}. For each 24-hour flaring rate [$\lambda$] flaring-rate uncertainties were calculated using the Poisson error [$\Delta \lambda = N^{-0.5}$] where $N$ represents the number of sunspot groups that underwent that evolution step. These appear as error bars in Figure~\ref{fig:zur_flare_nolimb} and ``$\pm$'' quantities on the rates given in Appendix~\ref{app_zur}. The maximum uncertainty associated with any flaring rate is therefore $\pm$\,1 flare per 24\,hours, corresponding to an evolution observed only once in the entire data set. Statistically significant rates therefore refer to those that exceed the magnitude of their Poisson error (\ie\ clearly separable from zero). It is worth mentioning that error bars are included in the panels of Figure~\ref{fig:zur_flare_nolimb} only when a non-zero rate was observed for that flaring level.

Initially focusing on the left column of Figure~\ref{fig:zur_flare_nolimb} (\ie\ $\geqslant$\,C1.0), the majority of flaring rates are deemed to be statistically significant and this is a consequence of high flare and evolution occurrence numbers. This means that these flaring rates indicate, with strong statistical certainty, the true rate associated with such evolution in Zurich class. For large-magnitude flares (\ie\ $\geqslant$\,M1.0 and $\geqslant$\,X1.0), the relative number of statistically significant rates becomes smaller. As the number of evolution occurrences (and hence Poisson rate uncertainties) do not change, this is due to there being markedly smaller numbers of M- and X-class flares. The only statistically significant $\geqslant$\,X1.0 flaring rate was that of the most complex Zurich F-class sunspot groups undergoing no evolution.

Similar analysis of flaring rates arising from Zurich class-specific evolution at the east and west limbs are provided in Appendix~\ref{app_zur}. These limb locations again display behaviour that diverges from that within $\pm$\,75\degree\ Heliographic longitude, in particular the overly high $\geqslant$\,C1.0 and $\geqslant$\,M1.0 flaring rates for sunspot groups starting as Zurich A-, B-, and C-class close to the east limb (Figure~\ref{fig:zur_flare_elimb}) and those starting as A-class close to the west limb (Figure~\ref{fig:zur_flare_wlimb}). These aberrant flaring rates close to the limb verify the mis-classification of Zurich classes identified previously in Sections~\ref{ssec:class_occ} and \ref{ssec:class_evol}. As a result, only flaring rates calculated from within $\pm$\,75\degree\ Heliographic longitude are presented for the penumbral class-specific evolution in Appendix~\ref{app_pen} and the compactness class-specific evolution in Appendix~\ref{app_comp}. Once again, both the penumbral and compactness flaring rates show qualitatively similar behaviour to that of the Zurich case, with increasingly higher flaring rates for greater upward evolution and opposite behaviour (\ie\ sequentially lower rates) for greater downward evolution.

    
\section{Conclusions}\label{sec:conc}

In this study we have examined the evolution of sunspot groups in terms of their McIntosh classification and their subsequent flaring rates. We have shown that the majority (\ie\ $\geqslant$\,60\,\%) of sunspot groups do not evolve on a 24-hour timescale for the McIntosh modified Zurich, penumbral, and compactness classes (\ie\ Figure~\ref{fig:hist_freq} and its equivalent in Appendices~\ref{app_pen} and \ref{app_comp}, respectively), with a secondary preference in overall evolution by $\pm$\,1 step in class. When examining limb-only locations (\ie\ those beyond $\pm$\,75\degree\ Heliographic longitude) we found that the overall evolution distributions show significant deviation from that observed on disk (\ie\ within $\pm$\,75\degree\ Heliographic longitude), with an inherent bias for evolution upward at the east limb and evolution downward at the west limb. This is a direct result of sunspot groups being mis-classified at both limbs due to foreshortening effects as sunspot groups rotate around the solar limb either into or out of view. This mis-classification manifests itself predominantly in the assignment of Zurich H-class (\ie\ unipolar with penumbra) whereby there is a tendency for over-classification of H-class at the east and west limbs. Taking this mis-classification into consideration, we therefore excluded both limbs from our main analysis of flaring rates and focus on only those sunspot groups within $\pm$\,75\degree\ Heliographic longitude.

The evolution of specific Zurich, penumbral, and compactness classes was examined and their resulting percentage occurrences analysed (\ie\ Figure~\ref{fig:zur_evol} and its equivalent in Appendices \ref{app_pen} and \ref{app_comp}, respectively). Again, it was found that sunspot groups predominantly do not evolve over 24\,hours and preferentially evolve by just $\pm$\,1 step in class. The Zurich occurrence evolution at the east and west limbs displays significant bias in terms of greater frequencies of upward evolution at the east limb and opposite behaviour (\ie\ downward evolution) at the west limb, reconfirming the mis-classification of Zurich classes at the limbs.

Class-specific evolution was examined further to calculate the subsequent 24-hour flaring rates associated with each evolution step. Increasingly higher flaring rates were observed in practically all starting classes for greater degrees of upward evolution in Zurich, penumbral, and compactness class (\ie\ Figure~\ref{fig:zur_flare_nolimb} and its equivalent in Appendices \ref{app_pen} and \ref{app_comp}, respectively), with opposite behaviour (\ie\ sequentially lower flaring rates) observed for greater downward evolution. For example, Figure~\ref{fig:zur_flare_nolimb} and Table~\ref{T-Zur_rates} show that sunspot groups which start as Zurich D-class and do not evolve yield a $\geqslant$\,C1.0 flaring rate of 0.68 
 flares per 24\,hours, while the rate for those that evolve upwards to E-class is 1.38 
 flares per 24\,hours and those evolving further upward to F-class is 2.67 
 flares per 24\,hours (\ie\ roughly double and quadruple, respectively, the rate of the no evolution case). In contrast, the flaring rate of sunspot groups that start as D-class and evolve downward to C-class is 0.21 
 flares per 24\,hours and those evolving further downward to B- or H-class is 0.08 or 0.06 flares per 24\,hours (\ie\ roughly a third and a tenth, respectively, the rate of the no-evolution case).

The evolution in McIntosh classification, specifically in the Zurich and compactness classes, act as a proxy for the emergence (upward evolution) or decay (downward evolution) of magnetic flux in a sunspot group. Our analysis therefore shows that flux emergence into a region produces a higher number of flares compared to the decay of flux. This result complements previous studies relating magnetic-flux emergence to the production of flares. \cite{Lee2012} showed that for the largest and most flare-productive McIntosh classifications, sunspot groups that were observed to increase in spot area over 24\,hours produced higher flaring rates than similarly classified groups that decreased in spot area. Our results agree very well with this and show that the McIntosh classification components can accurately characterize the growth of sunspot groups. 
 In conjunction with this, \cite{Schrijver2005} showed that flares $\geqslant$\,C1.0 were $\approx$2.4 times more frequent in active regions undergoing flux emergence that leads to the production of current systems and non-potential coronae than in near-potential regions. This mirrors our finding that sunspot groups increasing in Zurich, penumbral, or compactness class over 24-hour timescales have systematically higher flaring rates than those showing no change in these classes.

Some of the highest rates of flaring were observed for upward evolution from the larger, more complex Zurich classes -- \eg\ bipolar and large sunspot groups that start as Zurich D- and E-classes and evolve to F-class show a $\geqslant$\,C1.0 rate of 2.66\,$\pm$\,0.28 and 2.31\,$\pm$\,0.09 flares per 24\,hours, respectively. It was also found that increasingly complex Zurich classes produce higher flaring rates even when there is no evolution (\ie\ no flux emergence or decay) in a sunspot group over 24\,hours. This behaviour was observed throughout all starting Zurich classes (\ie\ A to F) and flaring magnitudes (\ie\ $\geqslant$\,C1.0, $\geqslant$\,M1.0, and $\geqslant$\,X1.0), indicating that flaring rates are correlated with the starting level of Zurich complexity as well as evolution through the three McIntosh classification components.

Finally, we calculated the associated uncertainty in our flaring rates using standard Poisson errors, in order to determine which of the rates are statistically significant (\ie\ clearly separable from zero). It was found that the majority of the evolution-dependent $\geqslant$\,C1.0 flaring rates are statistically significant -- a direct consequence of high numbers of both flares and evolution occurrence. As flare magnitude increases the flare occurrence drops significantly, leading to less statistically significant rates (\eg\ $\geqslant$\,X1.0 flaring rates are only significant for Zurich F-class groups that remain F-class, with a rate of 0.06\,$\pm$\,0.04 flares per 24\,hours). However, the same systematic behaviour of higher flaring rates for greater upward evolution (and lower rates for greater downward evolution) still persist for $\geqslant$\,M1.0 and $\geqslant$\,X1.0, despite the large uncertainties in these rates.

The evolution-dependent flaring rates presented here show potential for use in flare forecasting. Future work will focus on calculating evolution-dependent flaring probabilities under the assumption of Poisson statistics \citep{Gallagher2002}. The forecast performance of flaring rates determined here from Cycle 22 will be tested against data from Cycle 23 (\ie\ 1 August 1996 to 31 December 2010, inclusive), enabling direct comparison to the benchmark performance of the standard point-in-time (\ie\ not considering evolution) McIntosh--Poisson flare forecasting method presented in \citet{Bloomfield2012}.


 \begin{acks}
The authors thank Dr Chris Balch (NOAA/SWPC) for providing the data used in this research. AEM was supported by an Irish Research Council Government of Ireland Postgraduate Scholarship, while DSB was supported by the European Space Agency PRODEX Programme as well as the European Union's Horizon 2020 research and innovation programme under grant agreement No.~640216 (FLARECAST project).
 \end{acks}

%
%
\newpage
\appendix

\section{Zurich Class}\label{app_zur}

\begin{figure}[!ht]
\begin{center}
\includegraphics[width=0.95\textwidth]{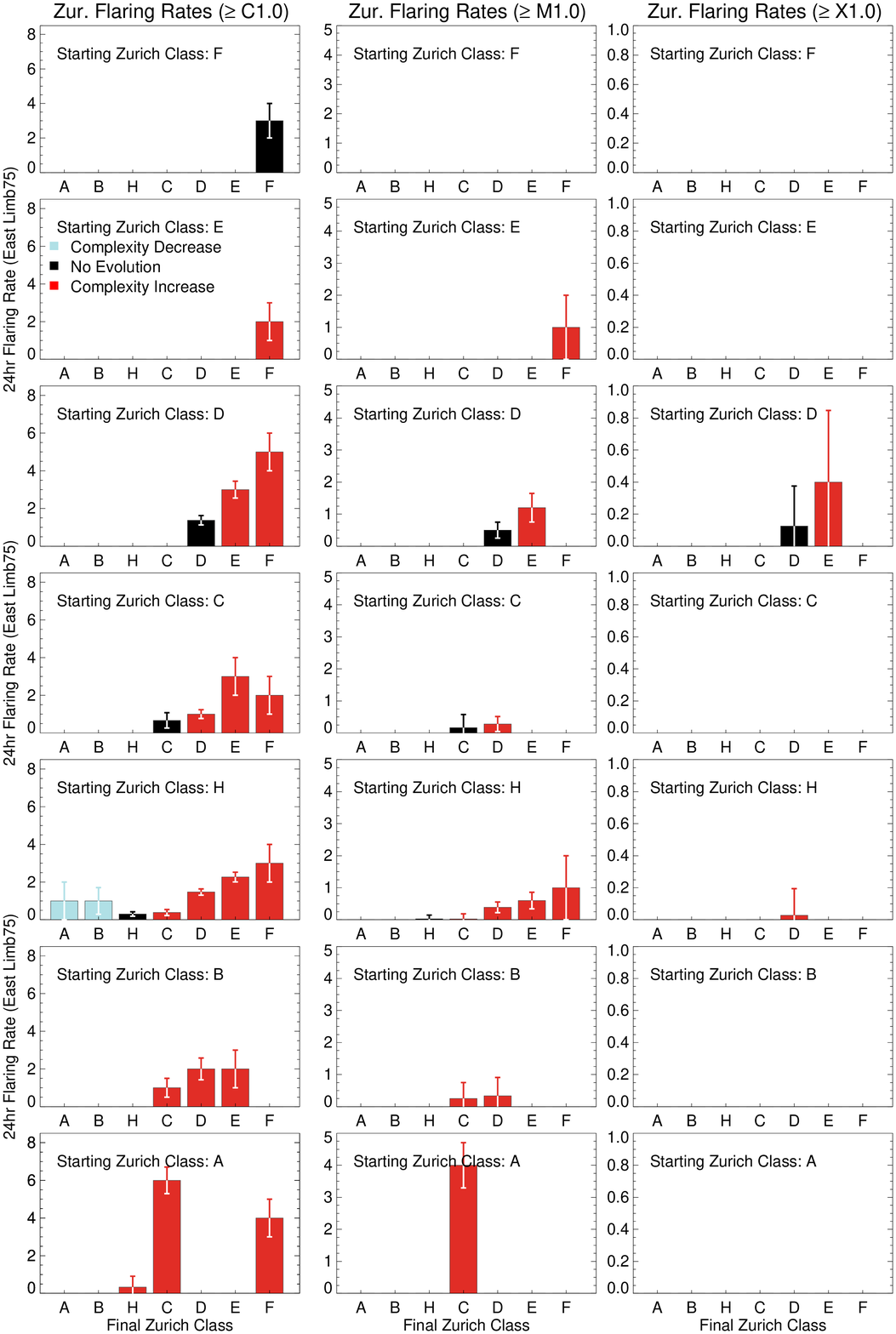}
\caption{Zurich evolution-dependent 24-hour flaring rates from groups at the east limb. Each column concerns different flaring levels: $\geqslant$\,C1.0 (left); $\geqslant$\,M1.0 (centre); $\geqslant$\,X1.0 (right). As in Figure~\ref{fig:zur_evol}, each row shows evolution from a different starting class and histogram bars are coloured by evolution: no change (black); upward evolution (red); downward evolution (blue).}
\label{fig:zur_flare_elimb}
\end{center}
\end{figure}

\begin{figure}[!ht]
\begin{center}
\includegraphics[width=0.95\textwidth]{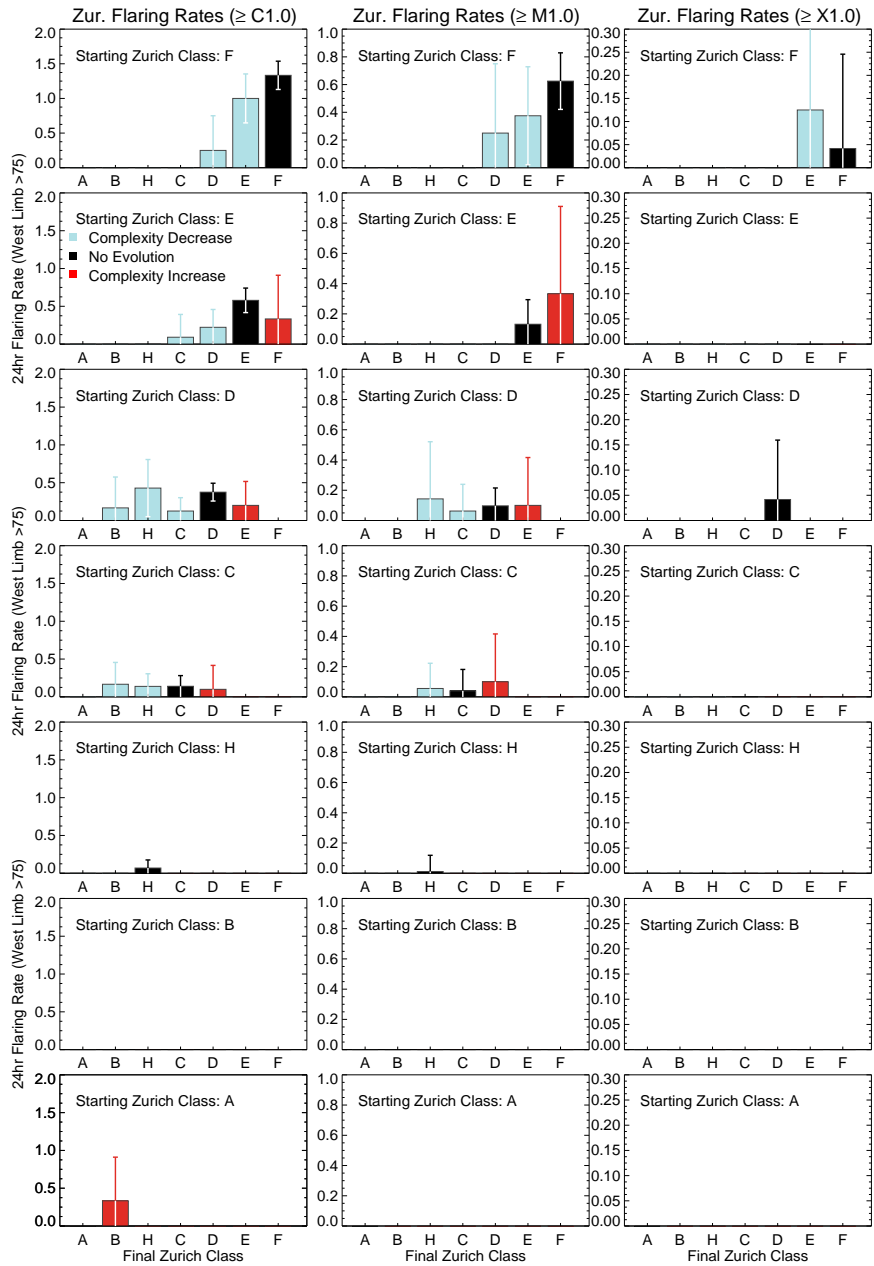}
\caption{Zurich evolution-dependent 24-hour flaring rates from groups at the west limb. Each column concerns different flaring levels: $\geqslant$\,C1.0 (left); $\geqslant$\,M1.0 (centre); $\geqslant$\,X1.0 (right). As in Figure~\ref{fig:zur_evol}, each row shows evolution from a different starting class and histogram bars are coloured by evolution: no change (black); upward evolution (red); downward evolution (blue).}
\label{fig:zur_flare_wlimb}
\end{center}
\end{figure}

Here we present accompanying tables and figures for the McIntosh Zurich class analysis. Evolution-dependent occurrence numbers and flaring rates of sunspot groups within $\pm$\,75\degree\ longitude are provided in Tables~\ref{T-Zur_occ} and \ref{T-Zur_rates}, respectively. These data correspond directly to that presented in Figures~\ref{fig:zur_evol} and \ref{fig:zur_flare_nolimb}, respectively. In addition, evolution-dependent flaring rates for the east and west limb regions (\ie\ $>$\,75\degree\ Heliographic longitude) are displayed in Figures~\ref{fig:zur_flare_elimb} and ~\ref{fig:zur_flare_wlimb}, respectively, for direct comparison to the ``within $\pm$\,75\degree\ longitude'' case presented in Figure~\ref{fig:zur_flare_nolimb}.

\begin{table}[!h]
\caption{Evolution-dependent McIntosh modified Zurich class occurrence numbers of sunspot groups within $\pm$\,75\degree\ Heliographic longitude}
\label{T-Zur_occ}
\begin{tabular}{lrrrrrrr}
\hline
Starting & \multicolumn{7}{c}{Ending class occurrence number}\\
class    & A    & B    & H    & C    & D    & E    & F       \\
\hline
F        &    0 &    1 &    0 &   11 &   13 &   79 &  451    \\
E        &    0 &    5 &    5 &  102 &  182 & 1059 &  111    \\
D        &    9 &   71 &   18 &  601 & 1839 &  338 &   12    \\
C        &  119 &  448 &  473 & 1767 &  562 &   44 &    8    \\
H        &  147 &   94 & 1378 &  362 &   55 &    9 &    2    \\
B        &  658 & 1254 &   34 &  367 &  166 &    0 &    0    \\
A        &  813 &  407 &   30 &  113 &   33 &    0 &    1    \\
\hline
\end{tabular}
\end{table}

\begin{sidewaystable}[!h]
\caption{Evolution-dependent McIntosh modified Zurich class flaring rates of sunspot groups within $\pm$\,75\degree\ Heliographic longitude }
\label{T-Zur_rates}
\begin{tabular}{lcccccccc}
\hline
Flaring           & Starting & \multicolumn{7}{c}{Ending class flaring rate [flares per 24\,h]} \\
level             & class    & A                 & B                 & H                 & C                 & D                 & E                 & F                \\
\hline
$\geqslant$\,C1.0 & F        & \ldots            & 0.00\,$\pm$\,1.00 & \ldots            & 0.09\,$\pm$\,0.30 & 1.62\,$\pm$\,0.28 & 1.52\,$\pm$\,0.11 & 2.43\,$\pm$\,0.05\\
\dotfill          & E        & \ldots            & 0.00\,$\pm$\,0.45 & 0.20\,$\pm$\,0.45 & 0.20\,$\pm$\,0.10 & 0.84\,$\pm$\,0.07 & 1.41\,$\pm$\,0.03 & 2.32\,$\pm$\,0.09\\
\dotfill          & D        & 0.00\,$\pm$\,0.33 & 0.08\,$\pm$\,0.12 & 0.06\,$\pm$\,0.24 & 0.21\,$\pm$\,0.04 & 0.68\,$\pm$\,0.02 & 1.38\,$\pm$\,0.05 & 2.67\,$\pm$\,0.29\\
\dotfill          & C        & 0.03\,$\pm$\,0.09 & 0.14\,$\pm$\,0.05 & 0.05\,$\pm$\,0.05 & 0.20\,$\pm$\,0.02 & 0.66\,$\pm$\,0.04 & 1.18\,$\pm$\,0.15 & 0.88\,$\pm$\,0.35\\
\dotfill          & H        & 0.01\,$\pm$\,0.08 & 0.09\,$\pm$\,0.10 & 0.05\,$\pm$\,0.03 & 0.23\,$\pm$\,0.05 & 0.89\,$\pm$\,0.13 & 1.44\,$\pm$\,0.33 & 3.50\,$\pm$\,0.71\\
\dotfill          & B        & 0.03\,$\pm$\,0.04 & 0.10\,$\pm$\,0.03 & 0.06\,$\pm$\,0.17 & 0.25\,$\pm$\,0.05 & 0.65\,$\pm$\,0.08 & \ldots            & \ldots           \\
\dotfill          & A        & 0.02\,$\pm$\,0.04 & 0.09\,$\pm$\,0.05 & 0.03\,$\pm$\,0.18 & 0.25\,$\pm$\,0.09 & 0.82\,$\pm$\,0.17 & \ldots            & 0.00\,$\pm$\,1.00\\
\hline
$\geqslant$\,M1.0 & F        & \ldots            & 0.00\,$\pm$\,1.00 & \ldots            & 0.00\,$\pm$\,0.30 & 0.08\,$\pm$\,0.28 & 0.33\,$\pm$\,0.11 & 0.65\,$\pm$\,0.05\\
\dotfill          & E        & \ldots            & 0.00\,$\pm$\,0.45 & 0.00\,$\pm$\,0.45 & 0.05\,$\pm$\,0.10 & 0.14\,$\pm$\,0.07 & 0.32\,$\pm$\,0.03 & 0.61\,$\pm$\,0.09\\
\dotfill          & D        & 0.00\,$\pm$\,0.33 & 0.01\,$\pm$\,0.12 & 0.00\,$\pm$\,0.24 & 0.02\,$\pm$\,0.04 & 0.11\,$\pm$\,0.02 & 0.30\,$\pm$\,0.05 & 0.83\,$\pm$\,0.29\\
\dotfill          & C        & 0.00\,$\pm$\,0.09 & 0.02\,$\pm$\,0.05 & 0.01\,$\pm$\,0.05 & 0.02\,$\pm$\,0.02 & 0.12\,$\pm$\,0.04 & 0.25\,$\pm$\,0.15 & 0.12\,$\pm$\,0.35\\
\dotfill          & H        & 0.00\,$\pm$\,0.08 & 0.00\,$\pm$\,0.10 & 0.01\,$\pm$\,0.03 & 0.03\,$\pm$\,0.05 & 0.22\,$\pm$\,0.13 & 0.22\,$\pm$\,0.33 & 0.00\,$\pm$\,0.71\\
\dotfill          & B        & 0.00\,$\pm$\,0.04 & 0.01\,$\pm$\,0.03 & 0.03\,$\pm$\,0.17 & 0.01\,$\pm$\,0.05 & 0.10\,$\pm$\,0.08 & \ldots            & \ldots           \\
\dotfill          & A        & 0.00\,$\pm$\,0.04 & 0.00\,$\pm$\,0.05 & 0.00\,$\pm$\,0.18 & 0.00\,$\pm$\,0.09 & 0.15\,$\pm$\,0.17 & \ldots            & 0.00\,$\pm$\,1.00\\
\hline
$\geqslant$\,X1.0 & F        & \ldots            & 0.00\,$\pm$\,1.00 & \ldots            & 0.00\,$\pm$\,0.30 & 0.00\,$\pm$\,0.28 & 0.03\,$\pm$\,0.11 & 0.07\,$\pm$\,0.05\\
\dotfill          & E        & \ldots            & 0.00\,$\pm$\,0.45 & 0.00\,$\pm$\,0.45 & 0.01\,$\pm$\,0.10 & 0.01\,$\pm$\,0.07 & 0.03\,$\pm$\,0.03 & 0.05\,$\pm$\,0.09\\
\dotfill          & D        & 0.00\,$\pm$\,0.33 & 0.00\,$\pm$\,0.12 & 0.00\,$\pm$\,0.24 & 0.00\,$\pm$\,0.04 & 0.01\,$\pm$\,0.02 & 0.03\,$\pm$\,0.05 & 0.00\,$\pm$\,0.29\\
\dotfill          & C        & 0.00\,$\pm$\,0.09 & 0.00\,$\pm$\,0.05 & 0.00\,$\pm$\,0.05 & 0.00\,$\pm$\,0.02 & 0.01\,$\pm$\,0.04 & 0.00\,$\pm$\,0.15 & 0.00\,$\pm$\,0.35\\
\dotfill          & H        & 0.00\,$\pm$\,0.08 & 0.00\,$\pm$\,0.10 & 0.00\,$\pm$\,0.03 & 0.00\,$\pm$\,0.05 & 0.02\,$\pm$\,0.13 & 0.00\,$\pm$\,0.33 & 0.00\,$\pm$\,0.71\\
\dotfill          & B        & 0.00\,$\pm$\,0.04 & 0.00\,$\pm$\,0.03 & 0.00\,$\pm$\,0.17 & 0.00\,$\pm$\,0.05 & 0.00\,$\pm$\,0.08 & \ldots            & \ldots           \\
\dotfill          & A        & 0.00\,$\pm$\,0.04 & 0.00\,$\pm$\,0.05 & 0.00\,$\pm$\,0.18 & 0.00\,$\pm$\,0.09 & 0.03\,$\pm$\,0.17 & \ldots            & 0.00\,$\pm$\,1.00\\
\hline
\end{tabular}
\end{sidewaystable}

\newpage
\clearpage

\newpage
\clearpage
\section{Penumbral Class}\label{app_pen}

\begin{figure}[!ht]
\begin{center}
\includegraphics[width=0.95\textwidth]{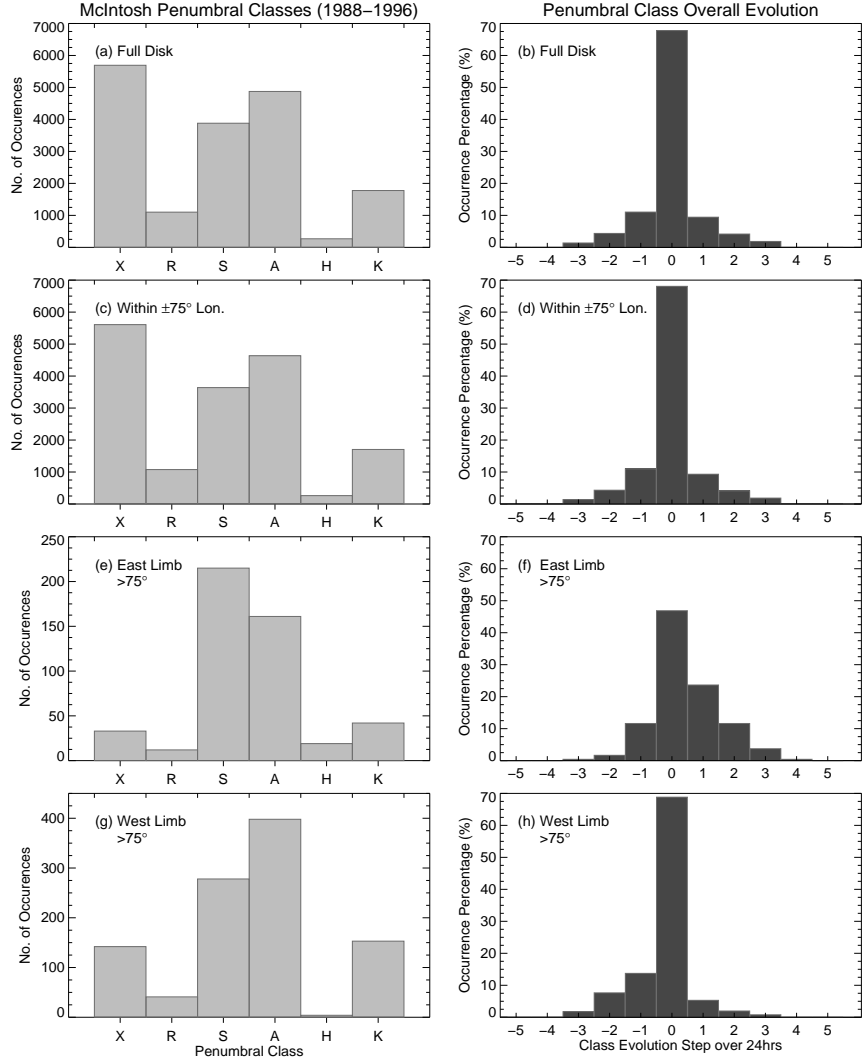}
\caption{Frequency histograms of each penumbral class in the McIntosh classification scheme (left column) and occurrence percentage histograms of their overall evolution steps on 24-hour timescales (right column). Each row presents data from different spatial locations on the Sun: full disk (panels a\,--\,b); within $\pm$\,75\degree\ Heliographic longitude (panels c\,--\,d); east limb (panels e\,--\,f); west limb (panels g\,--\,h). Positive evolution steps correspond to moving downwards through penumbral classes in Table~\ref{T-Pen}.}
\label{fig:pen_occ}
\end{center}
\end{figure}

\begin{figure}[!ht]
\begin{center}
\includegraphics[width=0.95\textwidth]{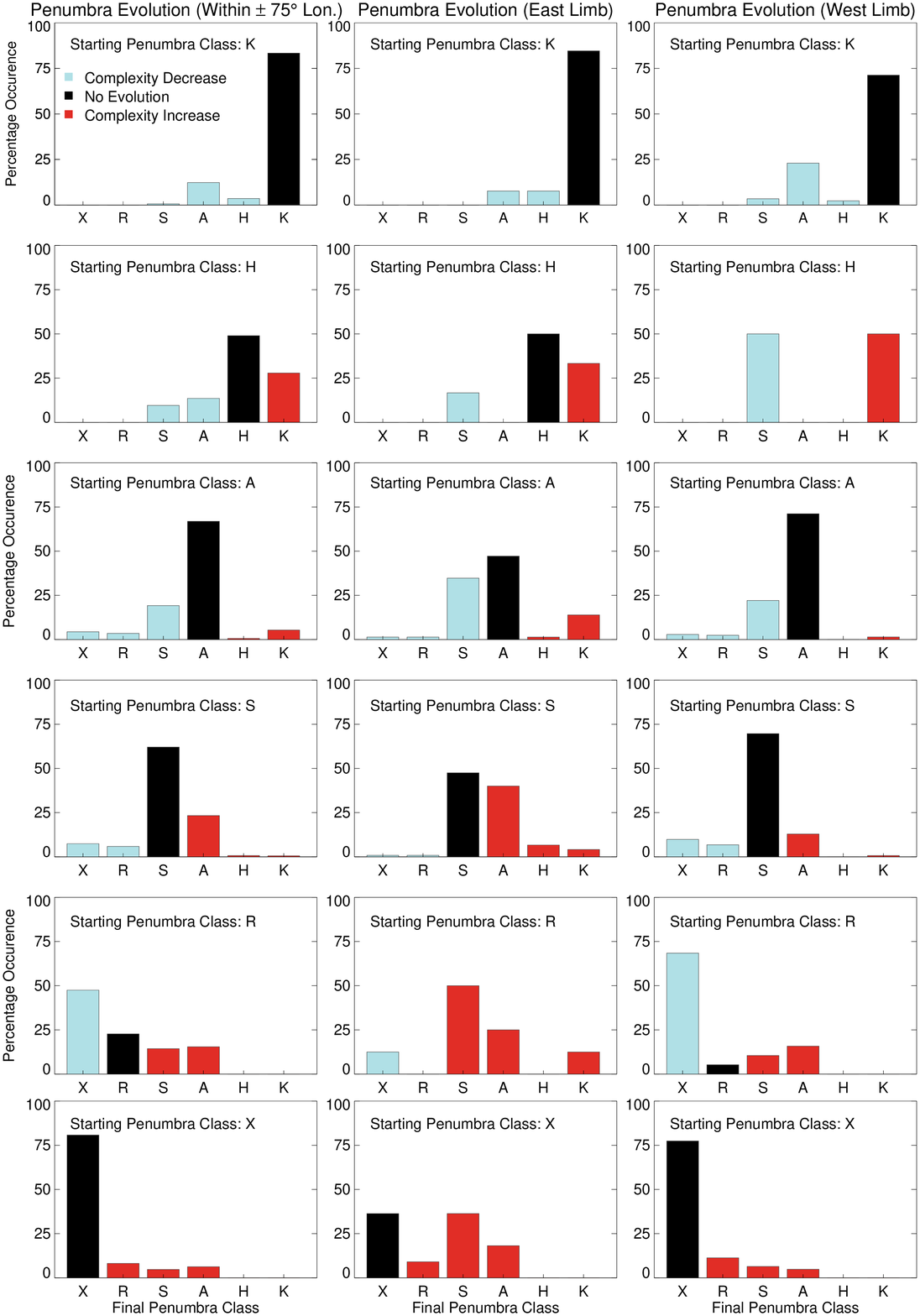}
\caption{Penumbral class 24-hour evolution histograms. Each column concerns different locations on the Sun: within $\pm$\,75\degree\ longitude (left); east limb (centre); west limb (right). Each row presents evolution from a different starting class, while bars give the percentage of that starting class coloured by evolution: no change (black); upward evolution (red); downward evolution (blue).}
\label{fig:pen_evol}
\end{center}
\end{figure}

\begin{figure}[!ht]
\begin{center}
\includegraphics[width=0.95\textwidth]{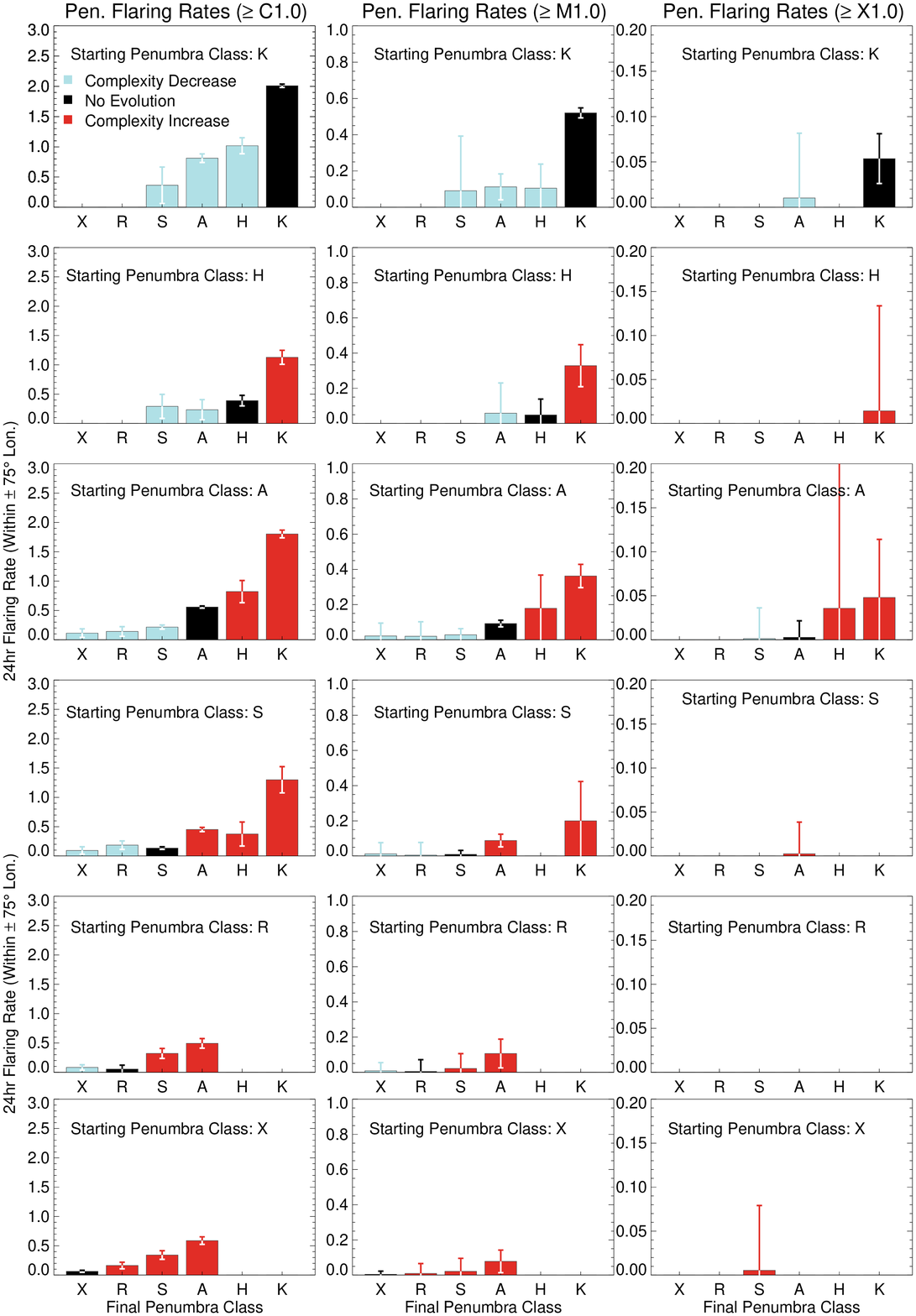}
\caption{Penumbral evolution-dependent 24-hour flaring rates from groups within $\pm$\,75\degree\ longitude. Each column concerns different flaring levels: $\geqslant$\,C1.0 (left); $\geqslant$\,M1.0 (centre); $\geqslant$\,X1.0 (right). As in Figure~\ref{fig:pen_evol}, each row shows evolution from a different starting class and histogram bars are coloured by evolution: no change (black); upward evolution (red); downward evolution (blue).}
\label{fig:pen_flare}
\end{center}
\end{figure}

Here we present equivalent tables and figures for the McIntosh penumbral class analysis. Frequency histograms of each penumbral class are shown in the left column of Figure~\ref{fig:pen_occ}, while overall evolution steps on a 24-hour timescale are given in the right column of Figure~\ref{fig:pen_occ} as percentage occurrence. Evolution-dependent occurrence numbers are provided in Table~\ref{T-Pen_occ} and graphically represented in Figure~\ref{fig:pen_evol} (equivalent to the Zurich class Table~\ref{T-Zur_occ} and Figure~\ref{fig:zur_evol}, respectively). Finally, flaring rates of sunspot groups within $\pm$\,75\degree\ longitude are provided in Table~\ref{T-Pen_rates} and graphically represented in Figure~\ref{fig:pen_flare} (equivalent to the Zurich class Table~\ref{T-Zur_rates} and Figure~\ref{fig:zur_flare_nolimb}, respectively).

\begin{table}[!h]
\caption{Evolution-dependent McIntosh penumbral class occurrence numbers of sunspot groups within $\pm$\,75\degree\ Heliographic longitude}
\label{T-Pen_occ}
\begin{tabular}{crrrrrr}
\hline
Starting & \multicolumn{6}{c}{Ending class occurrence number}\\
class    & X    & R    & S    & A    & H    & K              \\
\hline
K        &    0 &    0 &   11 &  196 &   57 & 1324           \\
H        &    0 &    0 &   24 &   34 &  123 &   70           \\
A        &  186 &  149 &  816 & 2843 &   28 &  229           \\
S        &  246 &  194 & 2056 &  772 &   24 &   20           \\
R        &  462 &  221 &  140 &  150 &    0 &    0           \\
X        & 3132 &  317 &  184 &  243 &    0 &    0           \\
\hline
\end{tabular}
\end{table}

\begin{sidewaystable}[!h]
\caption{Evolution-dependent McIntosh penumbral class flaring rates of sunspot groups within $\pm$\,75\degree\ Heliographic longitude}
\label{T-Pen_rates}
\begin{tabular}{lccccccc}
\hline
Flaring           & Starting & \multicolumn{6}{c}{Ending class flaring rate [flares per 24\,h]} \\
level             & class    & X                 & R                 & S                 & A                 & H                 & K                \\
\hline
$\geqslant$\,C1.0 & K        & \ldots            & \ldots            & 0.36\,$\pm$\,0.30 & 0.81\,$\pm$\,0.07 & 1.02\,$\pm$\,0.13 & 2.01\,$\pm$\,0.03\\
\dotfill          & H        & \ldots            & \ldots            & 0.29\,$\pm$\,0.20 & 0.24\,$\pm$\,0.17 & 0.39\,$\pm$\,0.09 & 1.13\,$\pm$\,0.12\\
\dotfill          & A        & 0.11\,$\pm$\,0.07 & 0.14\,$\pm$\,0.08 & 0.22\,$\pm$\,0.04 & 0.56\,$\pm$\,0.02 & 0.82\,$\pm$\,0.19 & 1.80\,$\pm$\,0.07\\
\dotfill          & S        & 0.09\,$\pm$\,0.06 & 0.19\,$\pm$\,0.07 & 0.14\,$\pm$\,0.02 & 0.45\,$\pm$\,0.04 & 0.38\,$\pm$\,0.20 & 1.30\,$\pm$\,0.22\\
\dotfill          & R        & 0.08\,$\pm$\,0.05 & 0.05\,$\pm$\,0.07 & 0.32\,$\pm$\,0.08 & 0.49\,$\pm$\,0.08 & \ldots            & \ldots           \\
\dotfill          & X        & 0.07\,$\pm$\,0.02 & 0.16\,$\pm$\,0.06 & 0.34\,$\pm$\,0.07 & 0.59\,$\pm$\,0.06 & \ldots            & \ldots           \\
\hline
$\geqslant$\,M1.0 & K        & \ldots            & \ldots            & 0.09\,$\pm$\,0.30 & 0.11\,$\pm$\,0.07 & 0.11\,$\pm$\,0.13 & 0.52\,$\pm$\,0.03\\
\dotfill          & H        & \ldots            & \ldots            & 0.00\,$\pm$\,0.20 & 0.06\,$\pm$\,0.17 & 0.05\,$\pm$\,0.09 & 0.33\,$\pm$\,0.12\\
\dotfill          & A        & 0.02\,$\pm$\,0.07 & 0.02\,$\pm$\,0.08 & 0.03\,$\pm$\,0.04 & 0.09\,$\pm$\,0.02 & 0.18\,$\pm$\,0.19 & 0.36\,$\pm$\,0.07\\
\dotfill          & S        & 0.01\,$\pm$\,0.06 & 0.01\,$\pm$\,0.07 & 0.01\,$\pm$\,0.02 & 0.09\,$\pm$\,0.04 & 0.00\,$\pm$\,0.20 & 0.20\,$\pm$\,0.22\\
\dotfill          & R        & 0.01\,$\pm$\,0.05 & 0.00\,$\pm$\,0.07 & 0.02\,$\pm$\,0.08 & 0.11\,$\pm$\,0.08 & \ldots            & \ldots           \\
\dotfill          & X        & 0.00\,$\pm$\,0.02 & 0.01\,$\pm$\,0.06 & 0.02\,$\pm$\,0.07 & 0.08\,$\pm$\,0.06 & \ldots            & \ldots           \\
\hline
$\geqslant$\,X1.0 & K        & \ldots            & \ldots            & 0.00\,$\pm$\,0.30 & 0.01\,$\pm$\,0.07 & 0.00\,$\pm$\,0.13 & 0.05\,$\pm$\,0.03\\
\dotfill          & H        & \ldots            & \ldots            & 0.00\,$\pm$\,0.20 & 0.00\,$\pm$\,0.17 & 0.00\,$\pm$\,0.09 & 0.01\,$\pm$\,0.12\\
\dotfill          & A        & 0.00\,$\pm$\,0.07 & 0.00\,$\pm$\,0.08 & 0.00\,$\pm$\,0.04 & 0.00\,$\pm$\,0.02 & 0.04\,$\pm$\,0.19 & 0.05\,$\pm$\,0.07\\
\dotfill          & S        & 0.00\,$\pm$\,0.06 & 0.00\,$\pm$\,0.07 & 0.00\,$\pm$\,0.02 & 0.00\,$\pm$\,0.04 & 0.00\,$\pm$\,0.20 & 0.00\,$\pm$\,0.22\\
\dotfill          & R        & 0.00\,$\pm$\,0.05 & 0.00\,$\pm$\,0.07 & 0.00\,$\pm$\,0.08 & 0.00\,$\pm$\,0.08 & \ldots            & \ldots           \\
\dotfill          & X        & 0.00\,$\pm$\,0.02 & 0.00\,$\pm$\,0.06 & 0.01\,$\pm$\,0.07 & 0.00\,$\pm$\,0.06 & \ldots            & \ldots           \\
\hline
\end{tabular}
\end{sidewaystable}

\newpage
\clearpage
\section{Compactness Class}\label{app_comp}

\begin{figure}[!ht]
\begin{center}
\includegraphics[width=0.95\textwidth]{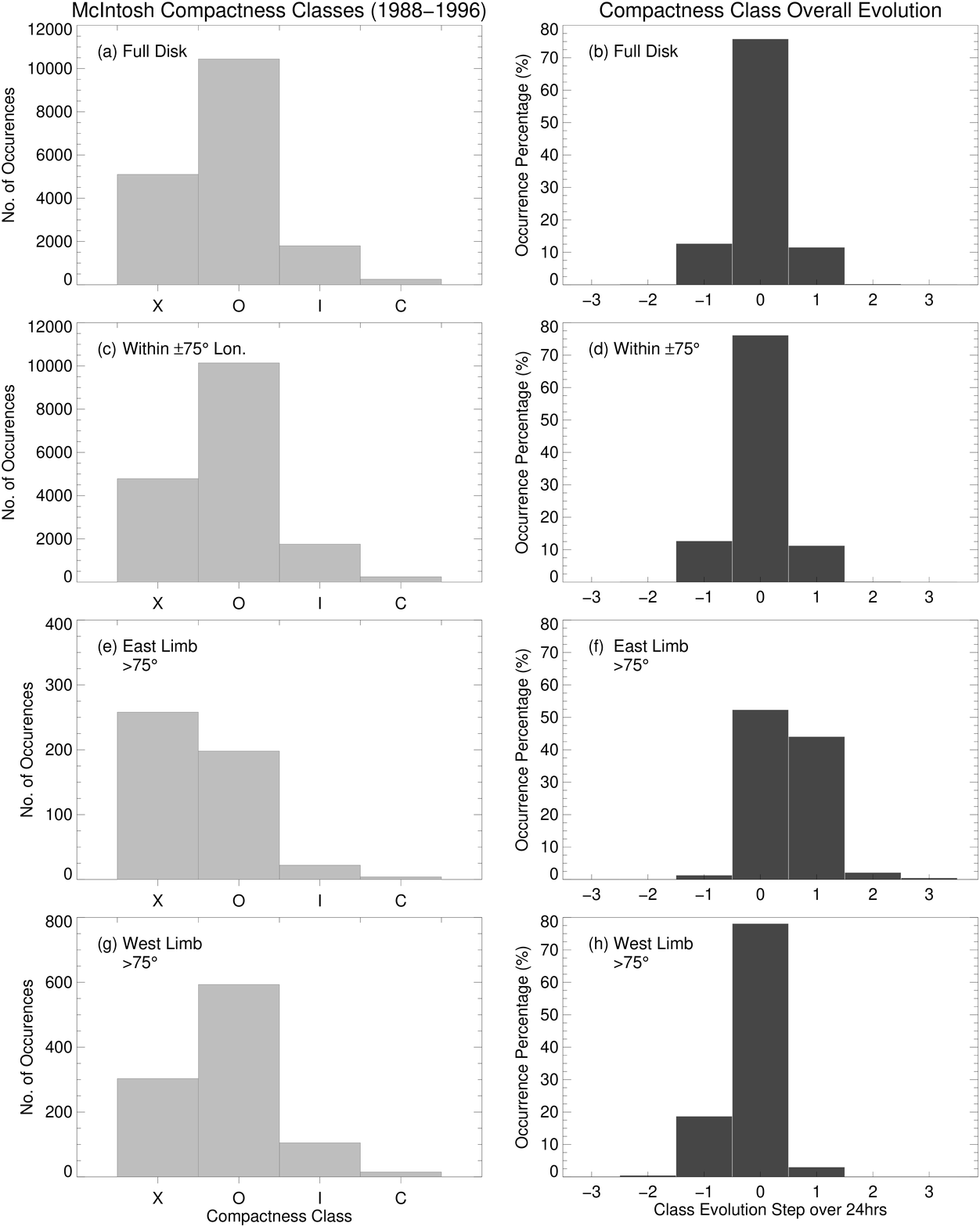}
\caption{Frequency histograms of each compactness class in the McIntosh classification scheme (left column) and occurrence percentage histograms of their overall evolution steps on 24-hour timescales (right column). Each row presents data from different spatial locations on the Sun: full disk (panels a\,--\,b); within $\pm$\,75\degree\ Heliographic longitude (panels c\,--\,d); east limb (panels e\,--\,f); west limb (panels g\,--\,h). Positive evolution steps correspond to moving downwards through compactness classes in Table~\ref{T-Comp}.}
\label{fig:comp_occ}
\end{center}
\end{figure}

\begin{figure}[!ht]
\begin{center}
\includegraphics[width=0.95\textwidth]{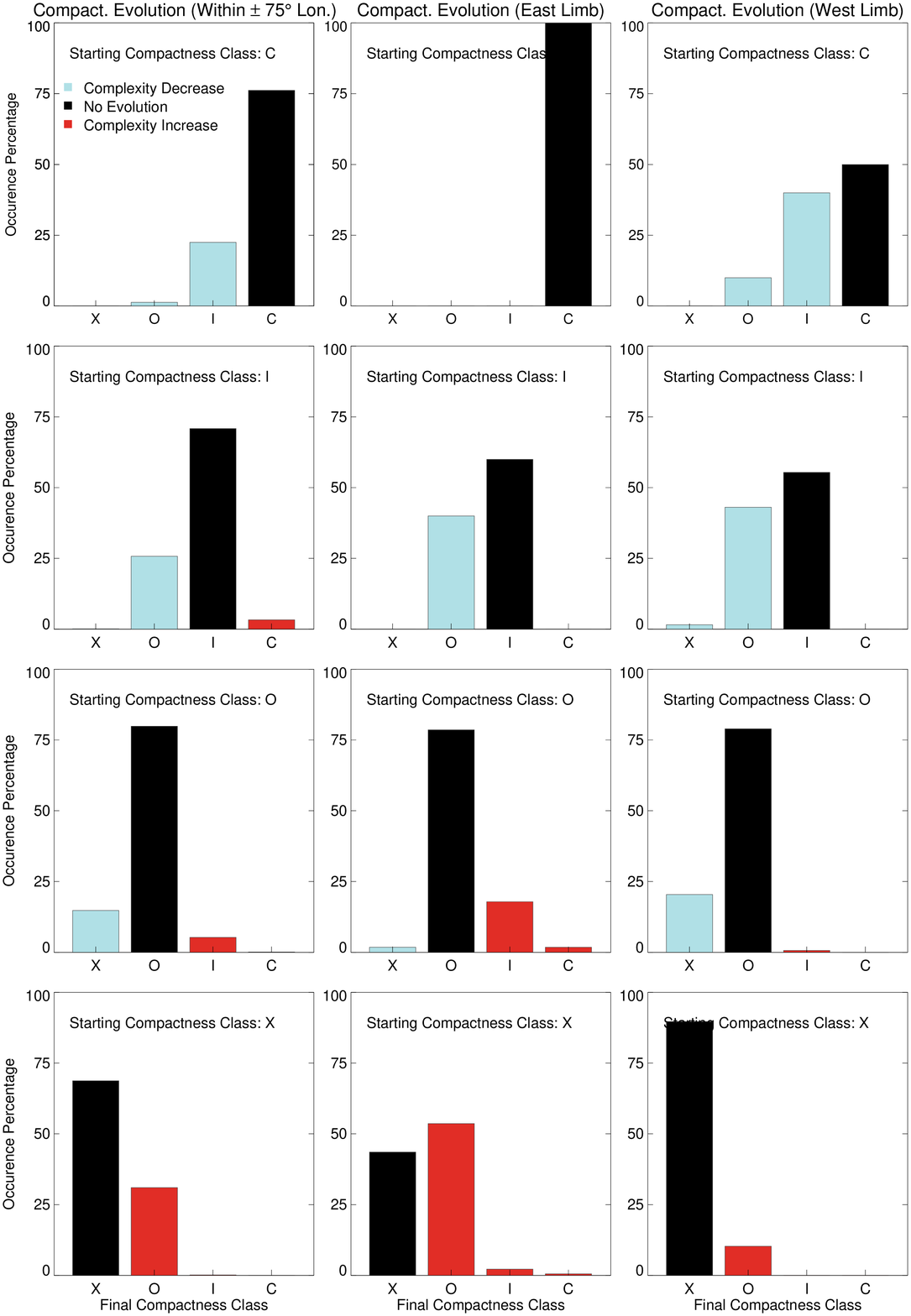}
\caption{Compactness class 24-hour evolution histograms. Each column concerns different locations on the Sun: within $\pm$\,75\degree\ longitude (left); east limb (centre); west limb (right). Each row presents evolution from a different starting class, while bars give the percentage of that starting class coloured by evolution: no change (black); upward evolution (red); downward evolution (blue).}
\label{fig:comp_evol}
\end{center}
\end{figure}

\begin{figure}[!ht]
\begin{center}
\includegraphics[width=0.95\textwidth]{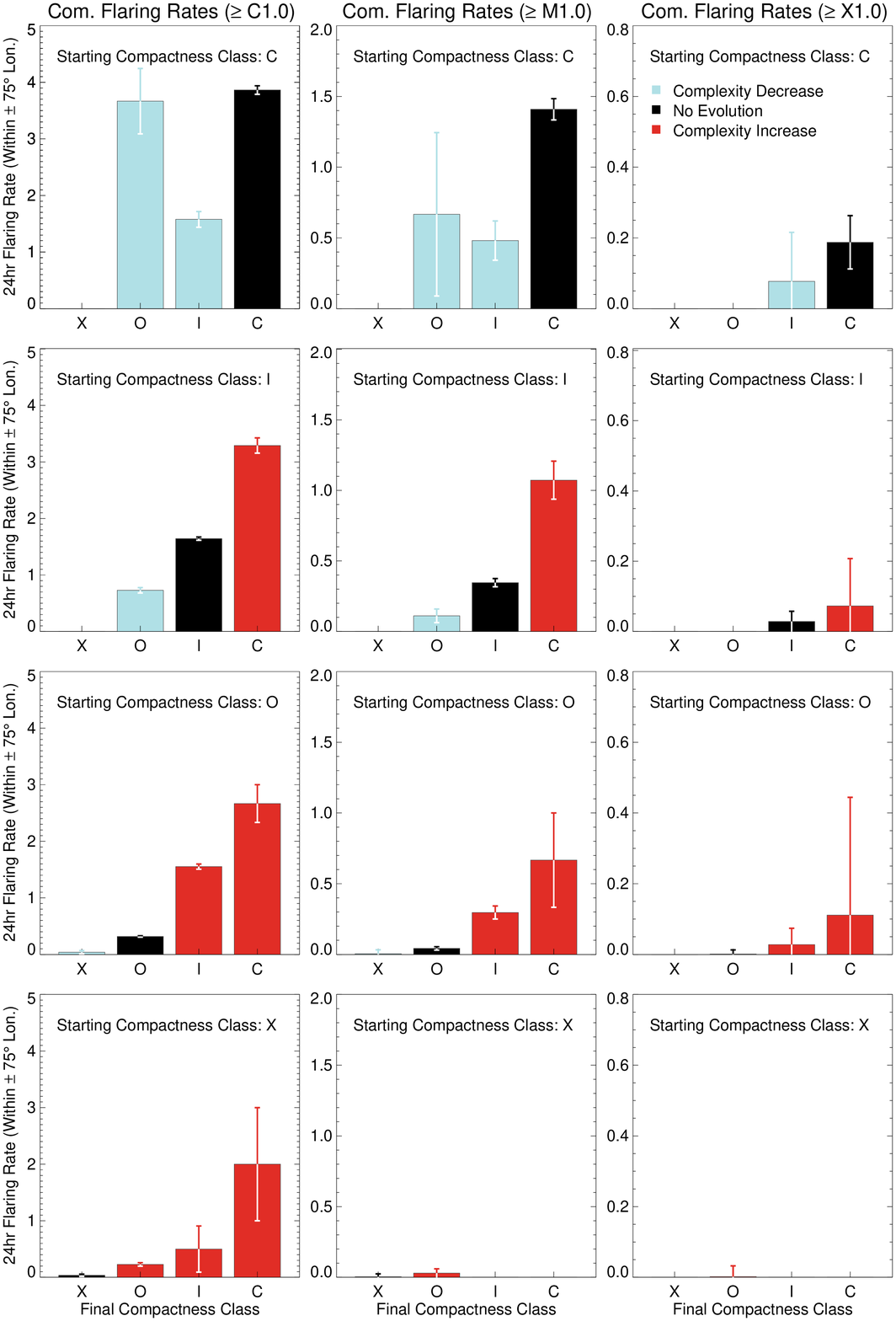}
\caption{Compactness evolution-dependent 24-hour flaring rates from groups within $\pm$\,75\degree\ longitude. Each column concerns different flaring levels: $\geqslant$\,C1.0 (left); $\geqslant$\,M1.0 (centre); $\geqslant$\,X1.0 (right). As in Figure~\ref{fig:comp_evol}, each row shows evolution from a different starting class and histogram bars are coloured by evolution: no change (black); upward evolution (red); downward evolution (blue).}
\label{fig:comp_flare}
\end{center}
\end{figure}

Here we present equivalent tables and figures for the McIntosh compactness class analysis. Frequency histograms of each compactness class are shown in the left column of Figure~\ref{fig:comp_occ}, while overall evolution steps on a 24-hour timescale are given in the right column of Figure~\ref{fig:comp_occ} as percentage occurrence. Evolution-dependent occurrence numbers are provided in Table~\ref{T-Comp_occ} and graphically represented in Figure~\ref{fig:comp_evol} (equivalent to the Zurich class Table~\ref{T-Zur_occ} and Figure~\ref{fig:zur_evol}, respectively). Finally, flaring rates of sunspot groups within $\pm$\,75\degree\ longitude are provided in Table~\ref{T-Comp_rates} and graphically represented in Figure~\ref{fig:comp_flare} (equivalent to the Zurich class Table~\ref{T-Zur_rates} and Figure~\ref{fig:zur_flare_nolimb}, respectively).

\begin{table}[!h]
\caption{Evolution-dependent McIntosh compactness class occurrence numbers of sunspot groups within $\pm$\,75\degree\ Heliographic longitude}
\label{T-Comp_occ}
\begin{tabular}{crrrr}
\hline
Starting & \multicolumn{4}{c}{Ending class occurrence number}\\
class    & X    & O    & I    & C                            \\
\hline
C        &    0 &    3 &   52 &  176                         \\
I        &    2 &  432 & 1190 &   55                         \\
O        & 1314 & 7104 &  469 &    9                         \\
X        & 2369 & 1069 &    6 &    1                         \\
\hline
\end{tabular}
\end{table}

\begin{table}[!h]
\caption{Evolution-dependent McIntosh compactness class flaring rates of sunspot groups within $\pm$\,75\degree\ Heliographic longitude}
\label{T-Comp_rates}
\begin{tabular}{lccccc}
\hline
Flaring           & Starting & \multicolumn{4}{c}{Ending class flaring rate [flares per 24\,h]}              \\
level             & class    & X                 & O                 & I                 & C                 \\
\hline
$\geqslant$\,C1.0 & C        & \ldots            & 3.67\,$\pm$\,0.58 & 1.58\,$\pm$\,0.14 & 3.86\,$\pm$\,0.08 \\
\dotfill          & I        & 0.00\,$\pm$\,0.71 & 0.73\,$\pm$\,0.05 & 1.64\,$\pm$\,0.03 & 3.29\,$\pm$\,0.13 \\
\dotfill          & O        & 0.04\,$\pm$\,0.03 & 0.32\,$\pm$\,0.01 & 1.55\,$\pm$\,0.05 & 2.67\,$\pm$\,0.33 \\
\dotfill          & X        & 0.04\,$\pm$\,0.02 & 0.23\,$\pm$\,0.03 & 0.50\,$\pm$\,0.41 & 2.00\,$\pm$\,1.00 \\
\hline
$\geqslant$\,M1.0 & C        & \ldots            & 0.67\,$\pm$\,0.58 & 0.48\,$\pm$\,0.14 & 1.41\,$\pm$\,0.08 \\
\dotfill          & I        & 0.00\,$\pm$\,0.71 & 0.11\,$\pm$\,0.05 & 0.35\,$\pm$\,0.03 & 1.07\,$\pm$\,0.13 \\
\dotfill          & O        & 0.01\,$\pm$\,0.03 & 0.04\,$\pm$\,0.01 & 0.30\,$\pm$\,0.05 & 0.67\,$\pm$\,0.33 \\
\dotfill          & X        & 0.00\,$\pm$\,0.02 & 0.03\,$\pm$\,0.03 & 0.00\,$\pm$\,0.41 & 0.00\,$\pm$\,1.00 \\
\hline
$\geqslant$\,X1.0 & C        & \ldots            & 0.00\,$\pm$\,0.58 & 0.08\,$\pm$\,0.14 & 0.19\,$\pm$\,0.08 \\
\dotfill          & I        & 0.00\,$\pm$\,0.71 & 0.00\,$\pm$\,0.05 & 0.03\,$\pm$\,0.03 & 0.07\,$\pm$\,0.13 \\
\dotfill          & O        & 0.00\,$\pm$\,0.03 & 0.00\,$\pm$\,0.01 & 0.03\,$\pm$\,0.05 & 0.11\,$\pm$\,0.33 \\
\dotfill          & X        & 0.00\,$\pm$\,0.02 & 0.00\,$\pm$\,0.03 & 0.00\,$\pm$\,0.41 & 0.00\,$\pm$\,1.00 \\
\hline
\end{tabular}
\end{table}

\clearpage
\bibliographystyle{spr-mp-sola}
\bibliography{evolution_paper}  

\end{article} 
\end{document}